\documentstyle[12pt,epsf,cite]{article}
\begin {document}
~\hspace*{5.0cm} ADP-AT-97-1 (Revised 3 September 1997)\\
~\hspace*{5.9cm} {\it Astroparticle Physics}, accepted\\

\begin{center}
{\large \bf Gamma-Rays and Neutrinos from Very Young Supernova Remnants}\\[1cm]
R.J. Protheroe, W. Bednarek$^*$, and Q. Luo\\Department of Physics and
Mathematical Physics\\ The University of Adelaide, Adelaide, Australia
5005\\$^*$permanent address: University of \L\'od\'z, 90-236 \L\'od\'z,\\ 
ul. Pomorska 149/153, Poland.\\[2cm]
\end{center}

\begin{center}
\underline{Abstract}\\
\end{center}

We consider the result of acceleration of heavy ions in the slot
gap potential of a very young pulsar with a hot polar cap.
Photodisintegration of the heavy ions in the radiation field of
the polar cap and pulsar surface gives rise to a flux of
energetic neutrons.  Some fraction of these neutrons interact
with target nuclei in the supernova shell to produce neutrino and
gamma-ray signals which should be observable from very young
supernova remnants in our galaxy for a range of pulsar
parameters.

\section{Introduction}

Young supernova remnants are probably sites of acceleration of
particles to high energies and, as a result of interactions, also
sources of high energy neutrinos and $\gamma$-rays
\cite{BerezPril78,Sato78,ghs87,GaisserHardingStanev89,Yamada87,BerezPtusk89}.
Following the occurrence of SN1987A it was expected that high
energy $\gamma$-rays would be detected from this object, but the
observations have so far been negative \cite
{boetal88a,boetal88b,boetal89,chetal88,ciaetal88,retal88,aetal93,vsetal93}.
Also, some models required the formation of a pulsar during the
supernova (SN) explosion which has not yet been
discovered. Nevertheless, SN1987A renewed interest in predicting
neutrino and $\gamma$-ray fluxes from supernova remnants (SNR)
for times soon after the explosion.  For example, the model in
which protons are accelerated of at the reverse shock which forms
in a relativistic wind from a pulsar as a result of its
confinement by the supernova envelope, proposed by Rees \&
Gunn~\cite{rg74}, has been considered in several papers
\cite{bg87,GaisserHardingStanev89,Yamada87,haretal91}.  In this
model, relativistic protons produce $\gamma$-rays and neutrinos
in collisions with the matter in the shell, and observable
$\gamma$-ray fluxes were predicted for SN 1987A if the power in
relativistic protons was $\sim 10^{39}$ erg s$^{-1}$. If the
pre-supernova star had a strong wind, neutrinos and $\gamma$-rays
could be produced in collisions of relativistic particles
accelerated by the supernova shock itself, with the surrounding
matter. For this scenario, Kirk et al.~\cite{kdb95} predicted
detectable fluxes of TeV $\gamma$-rays for SN~1993J, but not for
SN~1987A.

In the present paper, we consider yet another scenario in which
particle acceleration along open magnetic field lines may take
place in the pulsar magnetosphere if a neutron star with a high
magnetic field and a sufficiently short period forms during the
supernova explosion.  That acceleration of heavy ions by the Crab
pulsar may take place, has already been suggested, based on
modeling the observed synchrotron wisps in the Crab
nebula~\cite{hetal92,ga94}. In these papers it is proposed that a
significant fraction of the rotational kinetic energy loss of the
Crab pulsar is carried by ions which are injected with a rate
close to the Goldreich \& Julian current, and that these ions are
accelerated through at least 20\% of the total voltage available
on open field lines.

It is expected that soon after supernova explosions the surface
temperatures of newly formed neutron stars are much higher than
derived from recent observations of the thermal emission from the
surfaces of classical pulsars ~\cite{fetal92,hr93,ofz93}.  Just
after formation, a neutron star's surface temperature is expected
to be very high ($\sim 10^{11}$ K) but drop very rapidly, and
models of neutron star cooling predict temperatures of the order
of a few $10^6$---$10^7$ K during the first $\sim 10$ years after
the explosion~\cite{nt87,ogel91}.  However, the polar cap of a
neutron star can have a temperature $\sim 10^7$ K or higher, due
to the heating by electrons and $\gamma$-rays from cascades in
the pulsar magnetosphere~\cite{rs75,as79,hcn80}, and their
thermal emission depends only on the pulsar parameters.

Nuclei, probably mainly Fe nuclei, extracted from the neutron
star surface and accelerated to high Lorentz factors can be
photodisintegrated during propagation through the neutron star's
radiation field. Relativistic neutrons extracted from Fe nuclei
in this way propagate straight out and escape from the
magnetosphere, interacting in or passing through the surrounding
supernova remnant shell.  Those interacting in the shell produce
neutrino and $\gamma$-ray signals.  Those passing through the
shell decay into relativistic protons which contribute to the
pool of Galactic cosmic rays.

In Sect.~2 we investigate the photodisintegration of nuclei in
the radiation field of a pulsar.  In Sect.~3 we consider the
energetics of pulsars and the acceleration of particles in the
slot gap, as well as polar cap heating.  We then compute in
Sect.~4 the spectrum of neutrons escaping from the pulsar's
radiation field, and in Sect.~5 the expected light curves and
spectra of $\gamma$-rays and neutrinos.
The acceleration of Fe nuclei in the outer gap of classical
radio pulsars ~\cite{chr86a,chr86b}, and
photodisintegration of Fe on the nonthermal radiation of the outer gap
is considered in another paper \cite{BednarekProtheroe}.

\section{Photodisintegration of nuclei in pulsar thermal radiation}

The possibility of photodisintegration of Fe nuclei during their
propagation through thermal radiation from very young neutron
stars has been noted by Bednarek \& Karaku\l a~{\cite{bk95}.  The
importance of this process can be evaluated by calculating the
mean free path, $\lambda_A(\gamma_A,T)$, for relativistic nuclei
with mass number $A$ and Lorentz factor $\gamma_{A}$ for
photodisintegration, typically extracting one nucleon, in black body
radiation with temperature $T$ which is given by
\begin{eqnarray}
{1 \over \lambda_A(\gamma_A,T)} = \int \int 
{{\epsilon^{*}n(\epsilon,T) \sigma_A(\epsilon^{*})}
\over{2\gamma_{A}\epsilon}} \, {\rm d}\epsilon \, {\rm d}(\cos\theta),
\label{eq:mfp}
\end{eqnarray}
\noindent
where $n(\epsilon, T)$ is the photon number density per unit
photon energy at photon energy $\epsilon$ for black body
radiation at temperature $T$, $\sigma_A(\epsilon^{*})$ is the
cross section for photodissociation of a single nucleon from a
nucleus with mass number $A$ \cite{kt93}, $\epsilon^{*} =
\gamma_{A}\epsilon (1 - \beta_A \cos\theta)$ is the photon energy
in the rest frame of the nucleus, $\beta_A c$ is the velocity of
the nucleus, and $\theta$ is the angle between the direction of
the nucleus and a thermal photon.  In Fig.~\ref{fig1} we show,
for illustration, the mean free paths through isotropic black body
radiation with temperature $T$, for nuclei with various mass
numbers as a function of Lorentz factor. Note that the mean free
paths computed by us using Eq.~\ref{eq:mfp} differ slightly for
low $A$ nuclei in comparison with those given by Karaku\l a \&
Tkaczyk~\cite{kt93} because the cross section given by Eq.~2 of
their paper incorrectly separates the contributions from the
Breit-Wigner peak and the approximately constant cross section at
high energies.

The radiation field we adopt in the present work is anisotropic
and the intensity of radiation at a particular point in space has
the the following properties: in directions towards the neutrons
star's polar caps the intensity is equal to that of black body
radiation of temperature equal to that of the polar caps; in
directions towards other parts of the the neutrons star's surface
the intensity is equal to that of black body radiation of
temperature equal to that of the surface; and in all other
directions the intensity is zero.  The radiation field may differ
from this because of radiative transfer effects in the neutron
star's atmosphere due to the presence of a strong magnetic field,
and this may cause the intensity to depend on the angle to the
magnetic pole (see e.g.  ref.~\cite{Pavlov94}).  This effect may
modify pulse profiles of pulsed X-ray emission, but it would not
affect significantly the photodisintegration since the modulation
with angle varies by a factor less than 4 for photon energies
$\sim 1$ keV (typical thermal photon energies for $T=10^7$~K).

To check if photodisintegration of Fe nuclei propagating in the
anisotropic radiation field of a neutron star perpendicular to
its surface is important, we compute the average number of
neutrons extracted from Fe nuclei as a function of their Lorentz
factor $\gamma_{\rm Fe}$.  We have made the reasonable
approximation, that in every photodisintegration, only one
nucleon is extracted. In Fig.~\ref{fig2}(a) we show the average
number of neutrons extracted from Fe nuclei as a function of
Lorentz factor $\gamma_{\rm Fe}$ for different surface
temperatures.  It is clear that for a neutron star with a uniform
surface temperature, the temperature needs to be rather high,
above $\sim 10^7$~K, for the process to occur efficiently.  Such
surface temperatures are higher than expected except within a few
days of the explosion.  However, as we shall discuss in a later
section, neutron star polar caps may have very high temperatures
due to particle acceleration and cascading in the magnetosphere.
We show in Fig.~\ref{fig2}(b) the average number of neutrons
extracted from Fe nuclei as a function of Lorentz factor in the
radiation field of a heated polar cap of radius $10^5$~cm for
three temperatures, and we see that efficient photodisintegration
can take place in such a radiation field.

We should point out that the presence of a magnetic field can
have an effect on any process, including photodisintegration of
nuclei by thermal photons, due to a change in kinematics,
resonant effects, strong field particle wavefunctions, etc.  In
this paper, we negelect such effects which are probably small
since heavy nuclei are involved.  We shall, however, consider in
the next section the possibility of photodisintegration of nuclei
directly on the pulsar's magnetic field.

\section{Photodisintegration of nuclei in the pulsar's magnetic field}

The photodisintegration of iron nuclei may also occur because of
the presence of an extremely high magnetic field.  The component
of magnetic field perpendicular to the direction of motion of a
particle with velocity $\beta c$ and Lorentz factor $\gamma$,
$B_\perp(x, \gamma)$, at distance $x=\beta c t$ above the polar
cap will give rise to electric and magnetic fields at time
$t'=t/\gamma$ in the particle's rest frame which are, in SI units,
${E'}_\perp(t')= \gamma c B_\perp(x, \gamma)$ and $B'_\perp(t') =
\gamma B_\perp(x, \gamma)$ respectively.
These fields are characteristic of electromagnetic radiation, and
the time-integrated Poynting flux per unit angular frequency is
\begin{eqnarray}
{d F' \over d a d \omega'} = {2 \over \mu_0 c} | {E'}_\perp(\omega')|^2 ,
\end{eqnarray}
where ${E'}_\perp(\omega')$ is the Fourier transform of
${E'}_\perp(t')$, and the number of photons per unit area per
unit angular frequency is
\begin{eqnarray}
{d N' \over d a d \omega'} = {1 \over \hbar \omega'} 
{2 \over \mu_0 c} | {E'}_\perp(\omega')|^2 .
\end{eqnarray}
This method is referred to as the Weizaeker-Williams method of
virtual quanta (see Section 15.4 in ref.~\cite{Jackson}).

As a relativistic iron nucleus travels along a magnetic field
line with radius of curvature $\rho$ it will acquire a drift
velocity $v_d$ perpendicular to the field direction with order of
magnitude
\begin{eqnarray}
v_d \sim {c^2 \gamma \over \omega_B \rho},
\end{eqnarray}
where $\omega_B = 9.6 \times 10^7 ZB/A$~rad~s$^{-1}$ is the
non-relativistic cyclotron frequency and $B$ is in Tesla (see
e.g. ref.~\cite{Jackson}, p 587).  Thus, the perpendicular component of
magnetic field is given approximately by
\begin{eqnarray}
B_\perp(x,\gamma) \approx {v_d \over c} B(x).
\end{eqnarray}
We shall assume a dipole magnetic field
\begin{eqnarray}
B(x) = B_s (R/r)^3
\end{eqnarray}
where $r \approx R+x$, $R$ is the neutron star's radius, and we
take the radius of curvature to be that of the last open field
line \cite{rs75}
\begin{eqnarray}
\rho = 2 \left ( {r c \over \Omega} \right )^{1/2} 
\approx 1.4 \times 10^9 (r_8 P)^{1/2} \hspace{5mm} \rm cm
\end{eqnarray}
where $r_8 = r / 10^8$~cm, and $P$ is the period in s.

As a nucleus is extracted from the neutron star's surface and is
accelerated along an open field line, $B_\perp$ will initially
increase as $\gamma$ increases and then decrease as $\rho$
increases and $B$ decreases.  In the instantaneous rest frame of
the nucleus, $B'_\perp$ (and ${E'}_\perp$) will also increase
from zero to some maximum, and then decrease.  In the high
$\omega'$ limit, the the spectrum of virtual quanta will be
determined by the most rapid change in ${E'}_\perp=cB'_\perp$.
To see whether the presence of the pulsar's magnetic field is at
all important for photodisintegration, we consider an extreme
case: $B_\perp$ changes instantaneously from zero to some
constant high value,
\begin{eqnarray}
B'_\perp(x,\gamma) \approx \Theta(0) 
 \gamma_{\rm max} B_\perp(x=0,\gamma=\gamma_{\rm max}).
\end{eqnarray}
If enhanced
photodisintegration due to the magnetic field is unimportant for
this case, then we can safely neglect it in the present work.

For the case under consideration, the electric field in the
instantaneous rest frame varies as
\begin{eqnarray}
{E'}_\perp(t') \approx \Theta(0) {E'}_0
\end{eqnarray}
where 
\begin{eqnarray}
{E'}_0 = \gamma_{\rm max} c B_\perp(x=0,\gamma=\gamma_{\rm max}).
\end{eqnarray}
Then 
\begin{eqnarray}
{E'}_\perp(\omega') \approx i {{E'}_0 \over 2 \pi \omega'}
\end{eqnarray}
where $i=\sqrt{-1}$ and
\begin{eqnarray}
{d N' \over d a d \omega'} &=& {1 \over \hbar \omega'} 
{2 \over \mu_0 c} \left ( {{E'}_0 \over 2 \pi \omega'} \right )^2 \nonumber \\
 &=& {1 \over \pi \mu_0 c h} {E'}_0^2  \omega'^{-3}.
\end{eqnarray}
For a photodisintegration threshold of $\hbar \omega'_{\rm min}
\sim 30$~MeV ($\omega'_{\rm min} \approx 4.6 \times 10^{22}$
rad~s$^{-1}$) we obtain an optical depth for photodisintegration
on the magnetic field of
\begin{eqnarray}
\tau_{B} & \approx & \sigma_{\rm photo} \int_{\omega'_{\rm
  min}}^\infty {d N' \over d a d \omega'} d \omega' \nonumber \\
  & \approx & \sigma_{\rm photo} { 1 \over 3 \pi \mu_0 c h} {E'}_0^2
  {\omega_{\rm min}'}^{-2},
\end{eqnarray}
where $\sigma_{\rm photo} \sim 10^{-26}$~cm$^2$ is an
approximation to the energy-dependent photodisintegration cross
section of Fe nuclei above threshold.  Taking $B_s = 1.414 \times
10^{12}$ gauss, $P=5$~ms, $R=1.2 \times 10^6$~cm, and
$\gamma_{\rm max}=5 \times 10^7$, we obtain $E'_0 \approx 5
\times 10^{19}$ V~m$^{-1}$ giving $\tau_{B} \approx 5 \times
10^{-7}$.  Clearly, photodisintegration of Fe nuclei on the
pulsar's magnetic field is negligible for the presently assumed
pulsar parameters.

\section{Pulsar energetics and particle acceleration}

We are interested in the case of a very young pulsar which has
just been formed during a supernova explosion, and for a few
years after formation. It is assumed that pulsar acceleration can
operate very soon after the formation of a neutron star and can
tap into a substantial fraction of the spin-down power that is
not associated with gravitational radiation.  We approximate
this non-gravitational component of the spin-down power by the
magnetic dipole radiation power (erg s$^{-1}$) given by
\begin{eqnarray}
L_{\rm em}(\Omega, B) = B^2 R^6 \Omega^4 \sin^2i / 6 c^3,
\label{eq:power}
\end{eqnarray}
where $B$ is the surface magnetic field at the magnetic pole in
G, $R$ is the radius of the neutron star in cm, $\Omega=2 \pi/P$
where $P$ is the period in s, and $i$ is the angle between the
spin axis and rotation axis.  

The total magnetic dipole spin-down power should also include the
energy loss due to plasma flow (e.g. ref.~\cite{Nel88}).
However, pulsar spin-down through plasma winds is not well
understood for the case of $i \ne 0$.  Here, we use spin-down
through dipole radiation as a working hypothesis and $B \sin i$
is treated as one parameter (rather than having $B$ and $\sin i$
separately).

For the typical magnetic field of a classical pulsar $B \sim
10^{12}$~G, for acceleration of ions up to sufficiently high energies
for photodisintegration to take place we need to have a short
initial period.  Thus, in all the calculations we make in the
present paper we shall adopt $B \sin i =10^{12}$~G, and an
initial period of either $P_0=5$~ms or $P_0=10$~ms.

Equating $L_{\rm em}$ with the loss of rotational kinetic energy,
the period at time $t$ is given by the well-known formula,
\begin{eqnarray}
P^2(t) = 1.04 \times 10^{-15} t B_{12}^2 \sin^2 i + P_0^2,
\label{eq:p_evol}
\end{eqnarray}
where $B =10^{12}B_{12}$~G, $P_0$ is the initial pulsar period in
seconds, and we have used $R=1.2 \times 10^6$~cm and $I=1.4
\times 10^{45}$ g~cm$^2$.  Note that Eq.~\ref{eq:p_evol} neglects
other sources of loss of rotational kinetic energy, e.g. by
gravitational radiation or electromagnetic quadrupole radiation
(see the discussion by Manchester and Taylor
\cite{ManTaylor77book}).  The electromagnetic quadrupole
radiation which might be important in very young pulsars is very
uncertain and we shall neglect this.  However, gravitational
radiation losses are known to be important for very young pulsars
and the power in gravitational radiation (erg s$^{-1}$) is given
by
\begin{eqnarray}
L_{\rm gr}(\Omega, k) = - {32 \over 5} {G_N \over c^5} I^2 k^2 \Omega^6
\label{eq:gr_power}
\end{eqnarray}
where $G_N$ is the gravitational constant, $I$ is the moment of
inertia and $k$ is the ellipticity \cite{ShapiroTeukolsy}.  We
adopt $k=3 \times 10^{-4}$ which is required in the case of the
Crab pulsar to give a pulsar age equal to the time elapsed since
the observation of the Crab supernova in AD~1054 \cite{ShapiroTeukolsy}.

Starting with an initial pulsar period we have performed
numerical integrations over pulsar age to obtain the period as a
function of age for the case where both electromagnetic losses
and gravitational radiation losses are included.  We show in
Fig.~\ref{fig:spindown} the period as a function of age for $B
\sin i = 10^{12}$~G and $k=3 \times 10^{-4}$ for two
initial periods: $P_0=5$~ms and $P_0=10$~ms.

According to standard models of neutron star cooling, the surface
temperature decreases to $T_s \sim 10^7$ K at a few days after the
explosion, and later cools much more slowly, reaching $T_s \sim
4\times 10^6$ K in one year.  In our calculations, we
interpolate/extrapolate the results of Nomoto and Tsuruta
\cite{nt87} for which $\log(T_s/1$~K) is 6.99, 6.72, 6.58, and 6.55
at $\log(t/$1~y$)= -2$, $-0.94$, 0.08, and 1.11 respectively.

Models of high energy processes in pulsars predict polar cap
temperatures can be significantly higher because of the heating
caused the relativistic $e^{\pm}$ pairs and $\gamma$-rays created
in the gaps in the pulsar magnetosphere.  For the space charge
limited flow model~\cite{as79,ar81,ar83} a lower limit on the
polar cap temperature can be estimated
\begin{eqnarray}
T_c^{AS}\approx 2.6\times 10^5 P^{-19/32} B_{12}^{1/4} R_6^{5/16} 
\hspace{5mm} \rm K,
\label{eq:pc_t_moderate}
\end{eqnarray}
where $P$ is in seconds and $R = 10^6 R_6$ cm.  In deriving this
formula it is assumed that the heating rate (given by Eq.~75 in
Arons \& Scharlemann \cite{as79}) applies to the region of
the polar cap defined by the last open field line which
intersects the pulsar surface at $(r,\theta)=(R,\theta_c)$ where
\begin{eqnarray}
\theta_c \approx (\Omega R/c)^{1/2}
\end{eqnarray}
giving a polar cap radius of 
\begin{eqnarray}
r_c \approx (2\pi R^3/cP)^{1/2}.
\label{eq:pc_rad}
\end{eqnarray}

Arons \& Scharlemann~\cite{as79} note that the heating rate may
actually be even higher, as given by the model of Ruderman \&
Sutherland \cite{rs75}, if most of the electrons from $e^{\pm}$
pairs created in the polar gap can be reversed and fall on the
polar cap region. The Ruderman \& Sutherland heating rate (given
by Eq.~26 in their paper) defines an upper limit to the polar cap
temperature
\begin{eqnarray}
T_c^{RS}\approx 2.8\times 10^6 P^{-8/28} B_{12}^{6/28} R_6^{-17/28} 
\hspace{5mm} \rm K.  
\label{eq:pc_t_maximum}
\end{eqnarray}
Note that in both models the polar cap temperatures depends on
the pulsar parameters.  For very rapid pulsars with high magnetic
fields they can reach $\sim 10^7$ K or higher.  In the present
paper we shall consider three cases: (1) no polar cap heating
(the whole star has temperature $T_s$); (2) surface except for
polar cap has temperature $T_s$ and polar cap has the higher of
$T_s$ or $T_c^{AS}$ (Eq.~\ref{eq:pc_t_moderate}) referred to as
``moderate polar cap heating''; and (3) surface except for polar
cap has temperature $T_s$ and polar cap has the higher of $T_s$
or $T_c^{RS}$ (Eq.~\ref{eq:pc_t_maximum}) referred to as ``maximum
polar cap heating''.

For hot polar caps, thermionic emission of ions is allowed, and
because of the free supply of charges from the polar caps,
the rotation-induced potential should be space charge limited
(e.g. refs.~\cite{Michel74,Fawley77,as79}).
Whether ions or electrons are accelerated out along field lines
depends on the sign of the accelerating electric field.  
Here we consider only the case in which ions are accelerated.

It is usually assumed that the pulsar surface is composed largely
of highly ionized Fe although other compositions, e.g. He nuclei
\cite{RosenCameron72}, are also possible.  Here, we assume that
the all the outflowing ions are fully-ionized Fe
(e.g. \cite{Fawley77}).  Fe nuclei accelerated in the
magnetosphere may reach sufficiently high Lorentz factors
such that photodisintegration could occur efficiently. To gain an
impression about the pulsar parameters for which this
may be important, we assume particles are accelerated by the
electric field in the slot gap \cite{as79,ar81,ar83}.  In this
model the electric potential in the gap is given approximately by
\begin{eqnarray}
\Phi(x) \approx 1.13 \times 10^2 \theta_c^4\, B\, R\, g(x)\,
\eta(1-\eta^2)\,\sin\zeta\, \sin i \hspace{5mm} \rm V,
\label{Eq:slot_gap}
\end{eqnarray}
where $\eta$ is the ratio of the colatitude angle of the open
field lines at the surface to $\theta_c$, and $\zeta$ is the
azimuthal angle with respect to the magnetic pole. In the
equation, $g(x) \approx x^2$ for $x < 0.5 \theta_c$, and $g(x)
\approx \theta_c[(x + 1)^{0.5} - 1]$ for $x > 0.5\theta_c$, and
$x$ is the distance along the gap in units of $R$. In writing
Eq.~\ref{Eq:slot_gap}, we ignore the terms due to the complicated
geometry of the gap. The ion acceleration zone corresponds to
$i>\pi/2$.  We shall adopt $\eta = 0.1$ and $\zeta=3\pi/2$ in
our work.  This potential is shown in
Fig.~\ref{fig:slot_gap_potential} for $B\sin i=10^{12}$~G and for
periods equal to the two initial periods, i.e. $P=5$~ms and
$P=10$~ms.

\section{Spectrum of neutrons from photodisintegration of Fe nuclei}

The extraction of energetic neutrons from accelerated nuclei has
important consequences for the transport of energy from the
pulsar's vicinity to the nebula since they move balistically
through the magnetosphere and beyond the light cylinder. Their
Lorentz factors are approximately equal to the Lorentz factors of
the parent nuclei, and are sufficiently high that they typically
decay at large distances from the pulsar.  Initially the SNR
shell is opaque to neutrons so that most neutrons interact in the
dense shell.  Later, when the shell becomes more transparent to
neutrons, they typically move through the SNR shell and decay outside
it. Most of the energy of the neutrons is then taken by protons
which are trapped and isotropized by the ambient magnetic field,
and these protons then wait for target nuclei, i.e. the SNR
shell, to arrive.  When the SNR shell catches up with the
relativistic protons, production of $\gamma$-rays and neutrinos
occurs through hadronic collisions but, as we shall discuss in
Section~6, this contribution is negligible for the present
radiation field and electric potential.  The protons will also
contribute to the Galactic cosmic rays, although their
contribution is relatively small.

The dependence of the electric potential in the case of the slot
gap model as a function of distance (Eq.~\ref{Eq:slot_gap})
defines the Lorentz factor of Fe nuclei during their propagation
from the surface.  Using this potential we compute the average
number of neutrons extracted from Fe nuclei as a function of
pulsar age.  We show this in Fig.~\ref{fig:neutron_number}(a) for
$P_0=5$~ms for each of the three radiation field models: no polar
cap heating, moderate polar cap heating, and maximum polar cap
heating.  Note that with maximum polar cap heating the polar cap
temperature is not significantly higher than the stellar surface
for $t<2.2$~days, and with moderate polar cap heating the polar
cap temperature is not significantly higher than the stellar
surface for $t<18$~days.  In Fig.~\ref{fig:neutron_number}(b) we
show the average cumulative number of neutrons extracted from a
single Fe-nucleus during acceleration for $P=5$ ms as a function
of distance along the slot gap for the three radiation field
models.

Since at a particular distance, a neutron extracted from the
accelerated ion will have an energy determined by the slot gap
potential, we can now compute the energy spectra of neutrons
escaping from the pulsar radiation field per single Fe nucleus
accelerated.  In Fig.~\ref{fig:neutron_spectra} we show the
spectrum of neutrons extracted from a single Fe-nucleus during
acceleration, $N_n(E_n)$, for a pulsar with $P_0=5$~ms and $B\sin
i=10^{12}$~G, for the three radiation field models.  Results
are given for (a) $t=1.15$~days ($P=5$~ms), (b) $t=20.5$~days
($P=5.05$~ms), and (c) $t=1$~year ($P=5.69$~ms).  In
Fig.~\ref{fig:neutron_spectra_compare} we compare the spectrum of
neutrons at $t=1$~year produced for $P_0=5$~ms with that produced
with $P_0=10$~ms.

The differential production rate, $\dot{N_n}(E_n)$, of neutrons
extracted from Fe nuclei depends on the total rate of Fe
nuclei injected, $\dot{N}_{\rm Fe}$, and is given by
\begin{eqnarray}
\dot{N_n}(E_n) = \dot{N}_{\rm Fe} N_n(E_n) = 
{{\xi L_{\rm em}(\Omega, B)}\over{Ze\Phi_{\rm max}(\Omega,B)}} N_n(E_n)
\label{eq:neutron_rate} 
\end{eqnarray}
\noindent
where $L_{\rm em}(\Omega, B)$ is the magnetic dipole radiation
approximation to the total non-gravitational radiation spin-down
power (Eq.~\ref{eq:power}), $\Phi_{\rm max}(\Omega, B) =
\Phi(r_{\rm LC}/R)$ is the maximum acceleration potential
traversed by Fe nuclei given by Eq.~\ref{Eq:slot_gap} and $r_{\rm
LC}$ is the distance to the light cylinder up to which this
acceleration is assumed to occur, $Z = 26$ is the atomic number
of Fe, and $e$ unit electric charge. The parameter $\xi$ gives
the fraction of total pulsar power used to accelerate Fe nuclei
and has an upper-limit $\xi\approx \Phi_{\rm
max}/\Phi_0\approx0.29$ where $\Phi_0=0.5\theta^4_cBR$ is the
maximum potential drop across the polar cap (e.g. Luo et
al. \cite{LPB97}). Note that we write the rate of injection of
iron nuclei as the ratio of a fraction of the power, $\xi L_{\rm
em}$ to the maximum particle energy, $Ze\Phi_{\rm max}$ since the
ratio gives the same order-magnitude estimate of the injection
rate as that derived from Goldreich-Julian density.

\section{Gamma-ray and neutrino spectra }

We make the approximation that in shell-type SNR the shell has a
thickness which is much smaller than the radius of the shell,
$r_{SN}$.  The optical depth of the shell to nucleon-nucleon
collisions is then
\begin{eqnarray}
\tau_{pp} = M_{SN} \sigma_{pp}/ (4\pi r_{SN}^2 m_p) \approx 
3.5 \times 10^{-5} M_{10}/(\beta_{\rm SN} t_y)^2
\label{eq:tau_pp}
\end{eqnarray}
where $M_{SN} = 10 M_{10}$ M$_{\odot}$ is the mass ejected into
the shell during the explosion, $\sigma_{pp} \approx 3 \times
10^{-26}$ cm$^2$ is the proton--proton inelastic cross section,
$m_p$ is the proton mass, $t = t_{y}$ years is the time after the
explosion, and $v_{\rm SN}=\beta_{\rm SN}c$ is the shell
expansion velocity.  Initially the shell is optically thick.  For
example, if $M_{SN}=10M_\odot$ and $\beta_{\rm SN}=0.03$ the
optical depth is greater than 1 for $t<2$ months.  All the
results we shall present below are for $\beta_{\rm SN}=0.03$ and
$M_{SN} = 10$ M$_{\odot}$.

During the early phase most of the neutrons interact with matter
in the shell producing  $\gamma$-ray and neutrino fluxes.
These signals may be observed provided the $\gamma$-ray and
neutrino beams intersect the direction to the Earth, or sweep
across the Earth giving rise to a pulsed signal.  For a beaming
solid angle of $\Omega_b$ steradians, the  neutrino flux
may be calculated from
\begin{eqnarray}
F_\nu(E_\nu) \approx {\dot{N}_{\rm Fe} \over \Omega_b d^2}
[1 - \exp(-\tau_{pp})] \int N_n(E_n) P_{n \nu}^M(E_\nu,E_n) \, {\rm d}E_n
\end{eqnarray}
where $d$ is the distance to the SNR, and $P_{n
\nu}^M(E_\nu,E_n){\rm d}E_\nu$ is the number of neutrinos
produced with energies in the range $E_\nu$ to $(E_\nu + {\rm
d}E_\nu)$ (via pion production and subsequent decays) in multiple
nucleon-nucleon interactions of a nucleon of energy $E_n$.
Nucleon-nucleon interactions were treated as described in
\cite{Hillas81} and the pions produced were decayed using SIBYLL
\cite{Fle94}.  All the results given below are for
$\Omega_b=1$~sr, and for smaller beams the fluxes should
therefore be scaled up by a factor $\Omega_b^{-1}$.

In the case of $\gamma$-ray production, a significant fraction of
the $\gamma$-rays will interact by pair production with target
nuclei in the shell and will not be observed.  Also, during the
first few months, TeV $\gamma$-rays could interact with radiation
from the photosphere of the supernova \cite{prot87}.  By several
days after the explosion, the radius of the photosphere becomes
smaller than that of the SNR shell, where the neutrons interact,
and continues to decrease in radius relative to the shell.  Thus,
interactions with radiation from the photosphere is likely to be
a small effect here as the $\gamma$-rays will be traveling in
the directions of their parent neutrons, i.e. radially outwards
from the neutron star, and the angle between the directions of
the $\gamma$-rays and soft photons from the photosphere will be
small, reducing the interaction probability.  Concentrating on
pair production with matter, the optical depth of the shell is
determined by the mean free path for pair production which at
high energies is (9/7) of the radiation length, giving
$\tau_{\gamma p} \approx 0.7 \tau_{pp}$.  For a neutron
interacting at some fraction, $f$, of the way through the shell,
a $\gamma$-ray produced at that point would have a probability of
$\exp[-(1-f)\tau_{\gamma p}]$ of escaping from the shell.
Integrating over neutron interaction points ($\gamma$-ray
emission points) within the shell we arrive at the  gamma
ray flux at Earth,
\begin{eqnarray}
F_\gamma(E_\gamma) \approx {\dot{N}_{\rm Fe} \over \Omega_b d^2} \,
{\exp(-\tau_{pp})-\exp(-\tau_{\gamma p}) \over 
( \tau_{\gamma p} / \tau_{pp}) -1}
\int N_n(E_n) P_{n \gamma}^M(E_\gamma,E_n) \, {\rm d}E_n,
\end{eqnarray}
where $P_{n \gamma}^M(E_\gamma,E_n){\rm d}E_\gamma$ is the
number of $\gamma$-rays produced with energies in the range
$E_\gamma$ to $(E_\gamma + {\rm d}E_\gamma)$ (via pion production
and subsequent decays) by multiple nucleon-nucleon interactions
of a nucleon of energy $E_n$. Note that this neglects cascading
in the matter as a result of bremsstrahlung and subsequent pair
production, and so the  $\gamma$-ray flux at low energies will
be somewhat underestimated.  

Light curves for  neutrinos above 1 TeV, and $\gamma$-rays
above 100 MeV and 1 TeV, are shown for $B\sin i=10^{12}$~G in
Fig.~\ref{fig:prompt_lightcurve}(a) for $P_0=5$~ms and in
Fig.~\ref{fig:prompt_lightcurve}(b) for $P_0=10$~ms.
The results are shown here, and in the remaining figures, for a
distance of $d=10$ kpc.  In both cases, results are given for
three radiation field models (no polar cap heating, moderate polar
cap heating, and maximum polar cap heating), and the
pulsar period and the surface and polar cap temperatures vary
appropriately with time after the explosion.  In
Fig.~\ref{fig:prompt_spectrum}(a) we show the energy spectrum of
 $\gamma$-rays, and in Fig.~\ref{fig:prompt_spectrum}(b) the
energy spectrum of  neutrinos, both at $t=0.1$ year for
each initial period and radiation field model.  In
Fig.~\ref{fig:prompt_spectrum_ev} we show how the energy spectrum
of  neutrinos evolves with time for the case of
$P_0=5$~ms and the maximum polar cap heating model.
In both Figs.~\ref{fig:prompt_spectrum}(b) and \ref{fig:prompt_spectrum_ev}
we show the atmospheric neutrino background flux within $1^\circ$ and 
within $10^\circ$ of the source direction based on the intensity
calculated by Lipari \cite{Lipari}.

\section{Discussion and Conclusion}

As we shall see, we require fairly high acceleration efficiencies
in very young pulsars for observable fluxes of $\gamma$-rays and
neutrinos.  However, high energy $\gamma$-ray observations of
several pulsars appear to suggest an increase in $\gamma$-ray
efficiency with age.  This may be attributed to the relation
between acceleration efficiency for primary particles that start
the electron-positron pair cascade and the age, provided the
pulsed $\gamma$-ray emission from pulsars is from pair cascades.
This may be true for young pulsars with ages more than $\sim
10^3$ years.  For {\em very} young pulsars (a few months to a few
years) as discussed in this paper, thermal emission from the
polar caps or neutron star's surface may have an important effect
on particle acceleration.  For very young pulsars with strong
magnetic fields and hot polar caps, the efficiency may be higher
for younger pulsars.  This is because energy losses due to
resonant inverse Compton scattering are important and can prevent
primary electrons or positrons from starting a cascade in regions
near the polar caps \cite{LuoProtheroe97}.

We next discuss whether the signals we predict from young SNR are
observable with existing and planned experiments (see
e.g. ref.~\cite{semetal95} for a summary of $\gamma$-ray detector
thresholds).  The $\gamma$-ray light curve peaks at about 2
months after the explosion (see
Fig.~\ref{fig:prompt_lightcurve}).  The peak flux above 100~MeV
for a pulsar at 10~kpc with $B\sin i=10^{12}$~G ranges from $2
\times 10^{-10}$ cm$^{-2}$ s$^{-1}$ ($P_0=10$~ms, no polar cap
heating) to $4 \times 10^{-9}$ cm$^{-2}$ s$^{-1}$ ($P_0=5$~ms,
maximum polar cap heating) if $\xi \Omega_b^{-1}=1$~sr$^{-1}$.
The sensitivity of the EGRET detector on the Compton Gamma Ray
Observatory for 100 MeV $\gamma$-rays is $\sim 7\times 10^{-8}$
cm$^{-2}$ s$^{-1}$, and with this sensitivity $\gamma$-rays from
a Galactic source should easily be detected for several months if
a significant fraction of the electromagnetic power ($\xi \sim
1$) goes into accelerating heavy nuclei and the pulsar's
$\gamma$-ray beam sweeps across the Earth and has $\Omega_b <
0.07$~sr ($P_0=5$~ms, maximum polar cap heating) or $\Omega_b < 3
\times 10^{-3}$~sr ($P_0=10$~ms, no polar cap heating).  With
future 100 MeV $\gamma$-ray detectors, even lower detection
thresholds (e.g. $\sim 5\times 10^{-9}$ cm$^{-2}$ s$^{-1}$ for
GLAST) should enable detection even if $\Omega_b$ were larger.

The current generation of Cherenkov telescopes operating at TeV
energies (e.g. the Whipple Observatory, and the CANGAROO
telescope \cite{Cangaroo}) have thresholds of a $\sim 3 \times
10^{-12}$ cm$^{-2}$ s$^{-1}$, and GRANITE III (Whipple telescope
with new camera) is expected to have a threshold of $\sim
10^{-12}$ cm$^{-2}$ s$^{-1}$.  For Galactic sources, the
predicted TeV fluxes (see Fig.~\ref{fig:prompt_lightcurve}) are
well above the Whipple threshold and should easily be detected
even for the least optimistic case ($P_0=10$~ms, no polar cap
heating) provided $\xi^{-1}\Omega_b < 3.8$~sr, for the
most optimistic case ($P_0=5$~ms, maximum polar cap heating)
provided $\xi^{-1}\Omega_b < 75$~sr.  With the
sensitivity of current TeV $\gamma$-ray telescopes, it should be
possible to detect $\gamma$-rays from sources in the Magellanic
Clouds provided $\xi^{-1}\Omega_b$ is sufficiently high, and we shall
discuss below constraints placed on our model by the
non-detection of $\gamma$-rays from SN~1987A.

Available upper limits on the TeV $\gamma$-ray flux from SN~1987A
are as follows: $t\approx 10$ months (November 1987) ---
$2.3\times 10^{-11}$ cm$^{-2}$ s$^{-1}$ above 1 TeV (Raubenheimer
et al.~\cite{retal88}); $t \approx 1$ year --- $6.1\times
10^{-12}$ cm$^{-2}$ above 3 TeV (Bond et al.~\cite{boetal88b});
$t \approx 1$ year (January/February 1988) --- $1.6\times
10^{-10}$ cm$^{-2}$ s$^{-1}$ above 0.4 TeV (Chadwick et
al.~\cite{chetal88}).  Assuming a distance of 55~kpc, and if the
observer is inside the beam of energetic neutrons injected by the
pulsar, which will also define the $\gamma$-ray beam, then the
value of $\xi^{-1}\Omega_b$ can be constrained by the 1~TeV upper
limit mentioned above, giving $\xi^{-1}\Omega_b > 0.36$~sr for
the case of $P_0=5$~ms (see Fig.~\ref{fig:prompt_lightcurve}a).
However, we note that since no pulsar has yet been discovered
inside SN~1987A this could be interpreted as suggesting that we
are outside the beam of the neutron injection or that no pulsar
was formed.

The technique for constructing a large area (in excess of $10^4$
m$^2$) neutrino telescope has been known for more than a decade
\cite{BerZat77}.  In  pioneering work, the DUMAND Collaboration
developed techniques to instrument a large volume of water in a
deep ocean trench with strings of photomultipliers to detect
Cherenkov light from neutrino-induced muons.  Locations deep in
the ocean shield the detectors from cosmic ray muons.  The DUMAND
detector \cite{Learned91} was designed to be most sensitive to
neutrinos above about 1 TeV, and prototypes of other similar
experiments are already in operation such as that in Lake Baikal,
Siberia \cite{Baikal}, and NESTOR off the coast of Greece
\cite{NESTOR}.  An exciting recent development has been the
construction of a DUMAND type detector deep in the polar ice cap
at the South Pole.  This experiment called AMANDA uses the same
principle as DUMAND but takes advantage of excellent transparency
of the polar ice under extreme pressures and a stable environment
in which to embed the detectors \cite{AMANDA}.  These experiments
which operate typically above 1~TeV have the potential to be
expanded in the future to a detector on the 1 km$^3$ scale.  The
background to these experiments at 1 TeV is due primarily to
atmospheric neutrinos, but at higher energies there is an
uncertain background of prompt muons from charm production (see
e.g. Gaisser et al. \cite{GaisserHalzenStanev95} for a
discussion).  For recent reviews of the predicted intensity of
diffuse high energy neutrinos of astrophysical origin see
refs.~\cite{GaisserHalzenStanev95,ProtheroeEriceNu97}.

Referring to Figs.~\ref{fig:prompt_spectrum}(b) and
\ref{fig:prompt_spectrum_ev} we see that for good angular
resolution ($1^\circ$) the neutrino flux above 1 TeV is well
above the atmospheric background, and even for $10^\circ$
resolution the neutrino signal may be observable.  However,
observation of any signal would require the neutrino beam
(essentially the beam of the primary neutrons) to sweep across
the Earth.  For $1^\circ$ resolution and
$\xi\Omega_b^{-1}=1$~sr$^{-1}$, TeV neutrinos should be easily
observable if their flux remained constant at the level given in
Fig.~\ref{fig:prompt_spectrum}(b) for 0.1~y.  However, the
neutrino light curve drops rapidly at $t > 0.1$~y (see
Fig.~\ref{fig:prompt_lightcurve}(b)) and so very large neutrino
telescopes will be required.  With $10^\circ$ resolution,
detection of galactic sources at 1~TeV would be marginal even in
the most optimistic scenario except for the case of extremely
large detectors, e.g. km$^3$ \cite{Halzen96} which, with large
statistics, could see a weak signal significantly below the
atmospheric background.  Note, however, that the emission peaks
at energies $\sim 10$--100~TeV, and that telescopes sensitive in
this range should be able to detect a neutrino signal for the
$P_0=5$~ms case from a galactic source even if the resolution
were $10^\circ$.  A large detector with excellent angular
resolution would be required to observe very young SNR in the
Magellanic Clouds unless the emission were beamed strongly
towards the Earth.

We have neglected any signals arising from the protons extracted
from Fe-nuclei, and also signals from nuclei that were not
completely fragmented into nucleons in the radiation field of the
pulsar.  We assume these particles, being charged, will be
trapped in the central region of the SNR where the matter density
is expected to be relatively low, and would not contribute
significantly to the $\gamma$-ray and neutrino fluxes.  These
nuclei would be subject to adiabatic deceleration as the SNR
expands.  However, some would accumulate and eventually
contribute to the galactic cosmic rays towards the end of the
SNR's life.  Neutrons that do not interact promptly travel far
from the pulsar where they decay into protons which await the
arrival of target nuclei in the supernova shell, and then produce
a delayed isotropic neutrino and $\gamma$-ray signal.  We have
estimated these delayed neutrino and $\gamma$-ray fluxes for the
present radiation field models and slot gap potential.  The
$\gamma$-ray light curves peak at about 6 months with fluxes about
$10^{-3}$ of the peak flux from neutron collisions in the shell
for $\Omega_b=1$~sr.  Thus, unlike the case of $\gamma$-rays from
interactions of neutrons with the shell, the delayed fluxes are
unlikely to be observable.  The TeV neutrino light curve is of
the same magnitude as the TeV $\gamma$-ray light curve.  Those
protons from neutron decay outside the nebula also contribute to
the Galactic cosmic rays.  We estimate the total energy going
into cosmic rays, with typical energies of $10^6$~GeV, integrated
over the pulsar's age to be $\sim 10^{45}$~erg, which is much
less than the $\sim 10^{50}$~erg potentially available through
shock acceleration.

In the case of filled nebulae, the $\gamma$-ray and neutrino fluxes
could be higher than estimated here.  For example, $\tau_{pp}$
would be a factor of 3 higher than given by Eq.~\ref{eq:tau_pp}
in the case of uniform filling.  More importantly, cosmic rays
trapped inside the nebula would also continuously contribute to
the $\gamma$-ray and neutrino fluxes.  Such a model is discussed in
the context of the Crab Nebula in another paper
\cite{BednarekProtheroe}.

We have investigated the use of higher magnetic fields, e.g
$10^{13}$~G, and found that this has two main effects: (1) the
polar cap heating is increased resulting in more neutrons
extracted per Fe nucleus (but up to a maximum of 30), (2) the
maximum acceleration potential is increased.  The second effect
causes the peak of the $\gamma$-ray and neutrino emission to move
to higher energies, and the injection rate of Fe nuclei to
decrease (see Eq.~\ref{eq:neutron_rate}).  Overall, the net effect of
using extremely high magnetic fields is unexpectedly to reduce
the predicted fluxes to levels comparable to or lower than those
predicted here for typical pulsar magnetic fields.

We should point out that in this paper we have ignored the
fall-back of matter onto the neutron star.  If fall-back occurs,
which cannot be ruled out, there are two immediate consequences:
acceleration of lighter ions such as carbon nuclei may occur, or
acceleration is completely quenched.  In the present work we have
also ignored production of $\gamma$-ray lines.  The neutrons and
protons from the Fe photodisintegration will excite nuclei of C,
N, O, etc. in the supernova shell to make gamma-ray lines in the
1--10 MeV range.  These lines may be detectable after about 1
year, when the shell becomes optically thin to scattering and
absorption.  Calculation of the flux of these lines is however
beyond the scope of the present paper.

In Conclusion, we have considered a model for $\gamma$-ray and
neutrino emission by very young supernova remnants in which ions
are accelerated in the slot gap of a highly magnetized rapidly
spinning pulsar.  Energetic neutrons are extracted from the ions
by photodisintegration during interactions with thermal radiation
from the neutron star surface and hot polar caps.  Gamma
ray and neutrino signals are produced by energetic neutrons
interacting with target nuclei as they travel out through the SNR
shell.  For a limited range of pulsar parameters these signals
should be observable from very young SNR in our Galaxy with
existing and planned $\gamma$-ray and neutrino telescopes within
one or two years of the supernova explosion.  The emission peaks
above TeV energies and so ground-based optical Cherenkov
detectors are more sensitive than the lower energy satellite
telescopes.  Neutrino telescopes should also be able to detect
the predicted neutrino signals which peak at $\sim 10$ TeV
energies. 

\section*{Acknowledgements}

We thank J\"{o}rg Rachen for suggesting the use of the
Weizaeker-Williams method when considering photdisintegration by
a magnetic field.  W.B. thanks the Department of Physics and
Mathematical Physics at the University of Adelaide for
hospitality during his visit. Q.L.  acknowledges receipt of an
Australian Research Council (ARC) Postdoctoral Fellowship.  This
research is supported by a grant from the ARC.

\newpage

\begin{figure}[htb]
\vspace{16cm}
\includegraphics{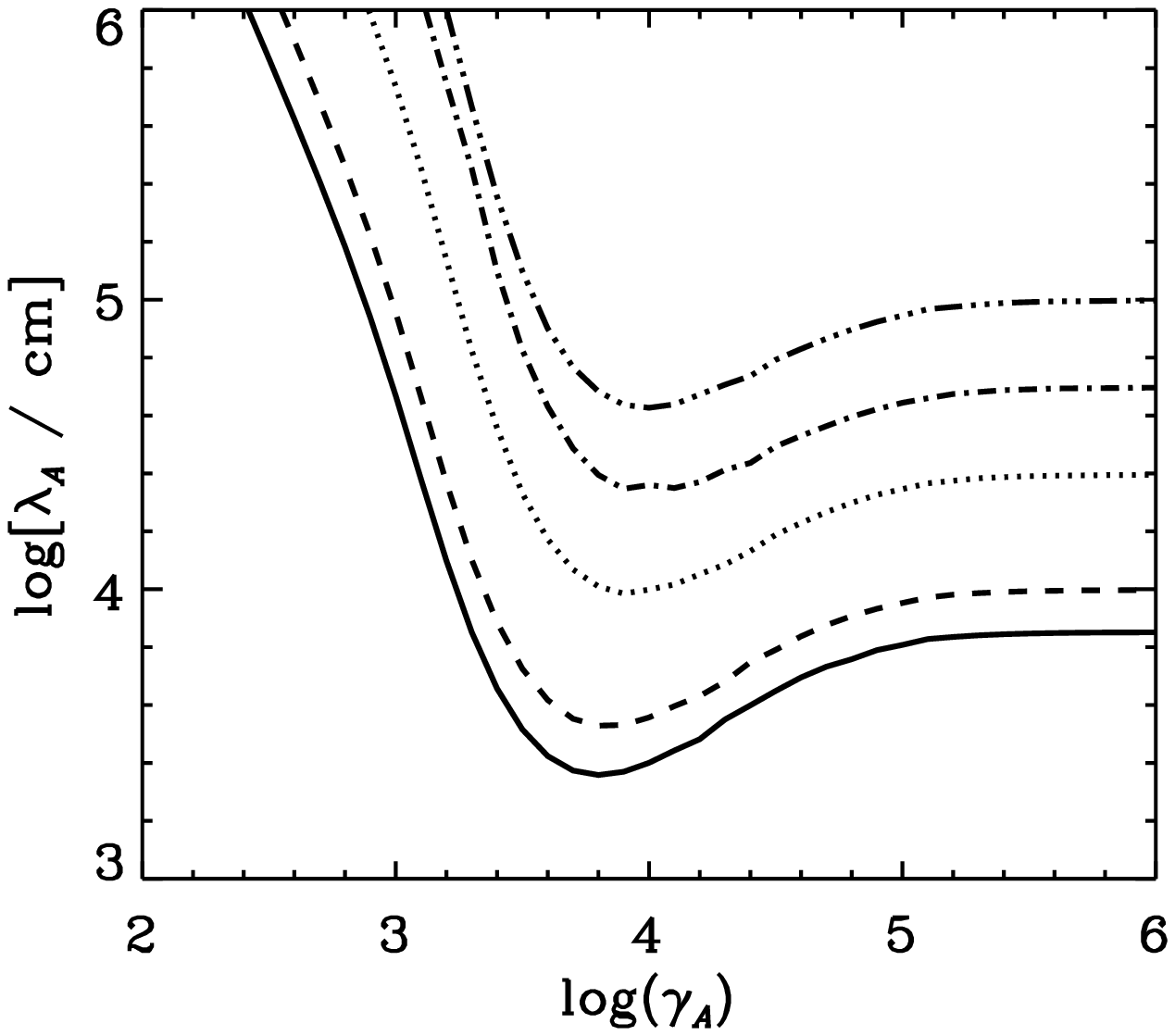}
\caption{Mean free path for photodisintegration of nuclei in
isotropic black body radiation at temperature $T=10^7$~K as a
function of Lorentz factor, $\gamma_A$, for various mass numbers
$A$: 4 (top curve), 8, 16, 40, and 56 (bottom curve).  }
\label{fig1}
\end{figure}

\begin{figure}[htb]
\vspace{16cm}
\includegraphics{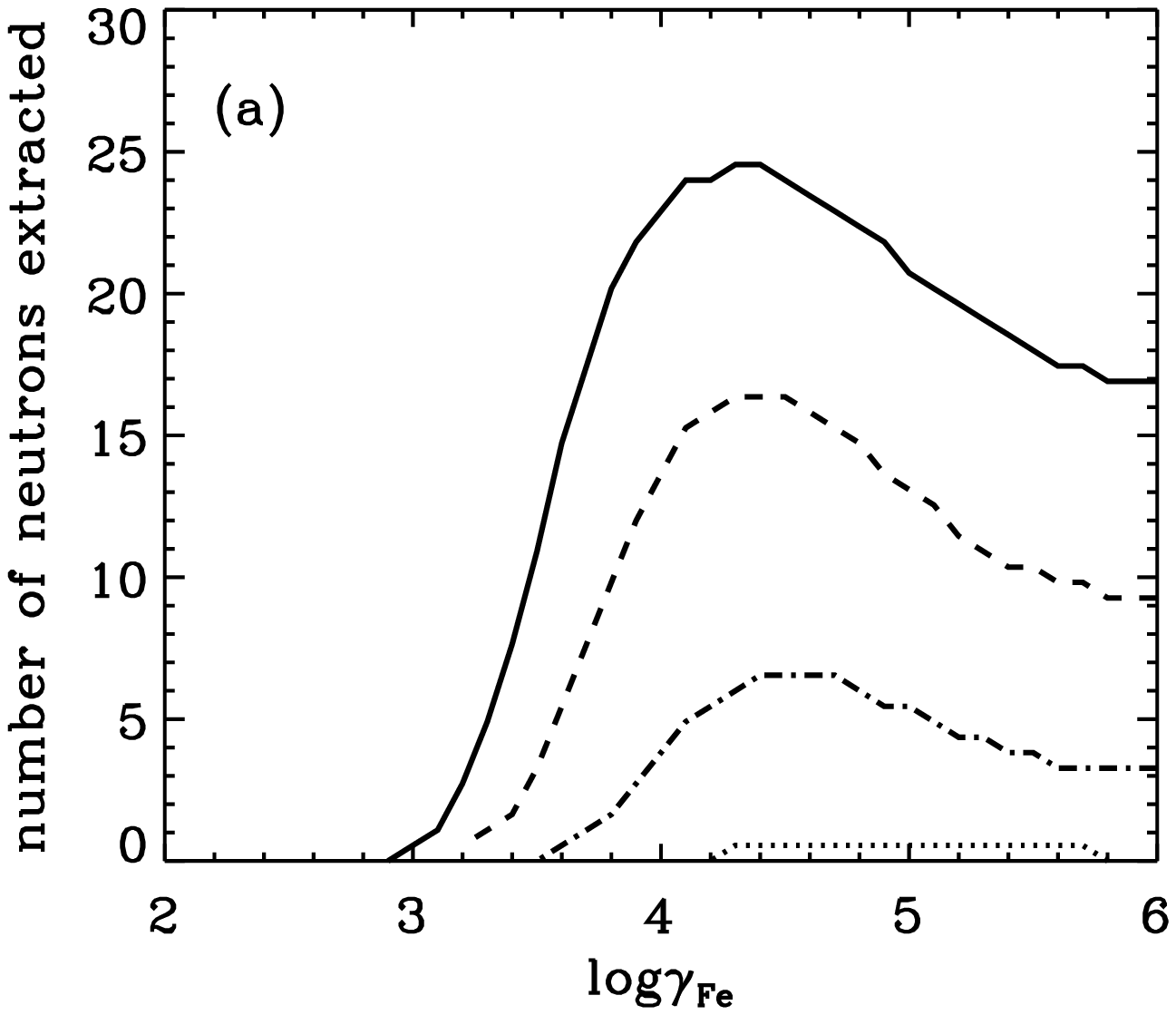}
\includegraphics{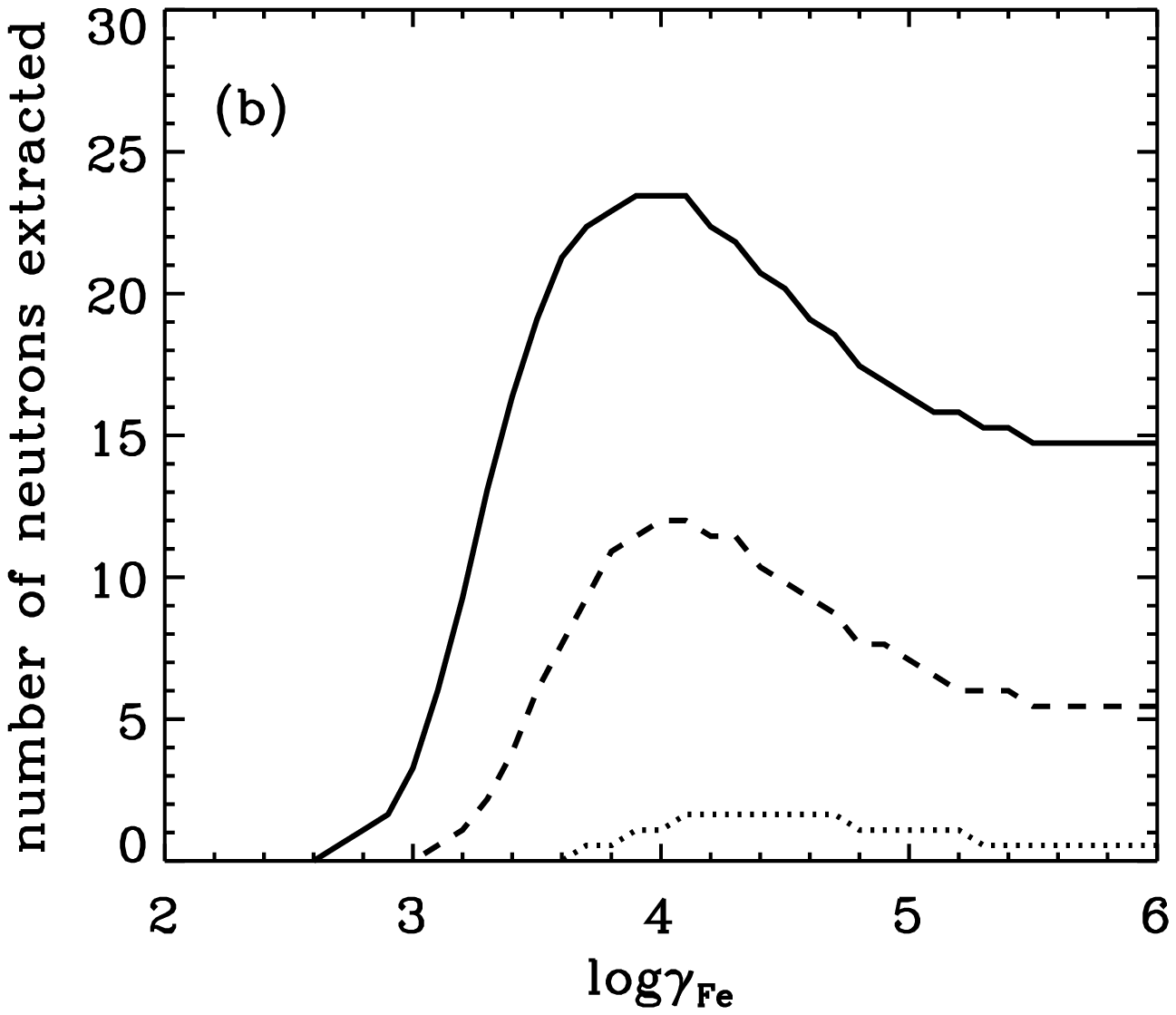}
\caption{(a) Average number of neutrons extracted from a single
Fe-nucleus during propagation along the neutron star's magnetic
axis, from its surface at the pole to infinity, as a function of
Lorentz factor.  Neutron star spherical surface of radius $1.2
\times 10^6$~cm at constant temperature $T=5 \times 10^6$~K
(bottom curve), $10^7$~K, $1.5 \times 10^7$~K, and $2 \times
10^7$~K (top curve).  (b) Average number of neutrons extracted
from a single Fe-nucleus during propagation along the neutron
star's magnetic axis, from its surface at the pole to infinity,
as a function of Lorentz factor, for field of heated polar cap of
radius $10^5$ cm with temperature $T=10^7$~K (bottom curve), $2
\times 10^7$~K, and $3 \times 10^7$~K (top curve).  }
\label{fig2}
\end{figure}

\begin{figure}[htb]
\vspace{16cm}
\includegraphics{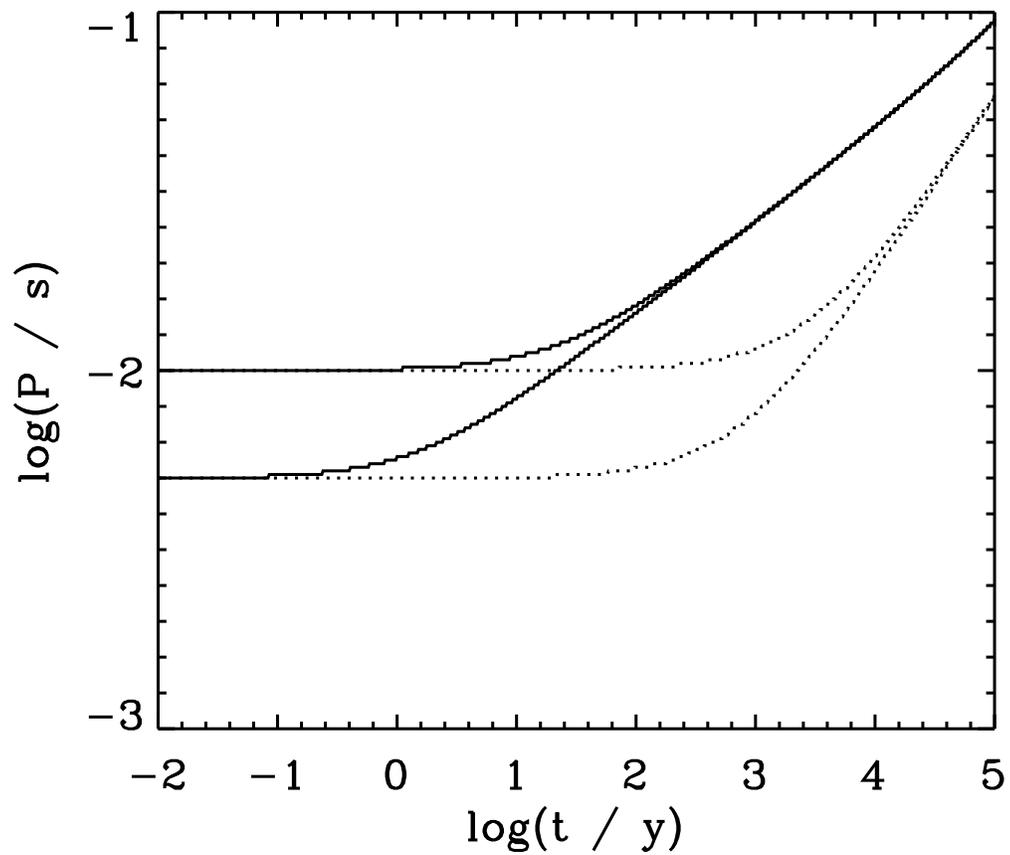}
\caption{Evolution of pulsar period for $B \sin i = 10^{12}$~G
for $P_0=5$~ms (lower curves) and $P_0=10$~ms (upper
curves).  Dotted curves are for electromagnetic radiation losses
only, and solid curves include the effect of gravitational
radiation losses. }
\label{fig:spindown}
\end{figure}

\begin{figure}[htb]
\vspace{16cm}
\includegraphics{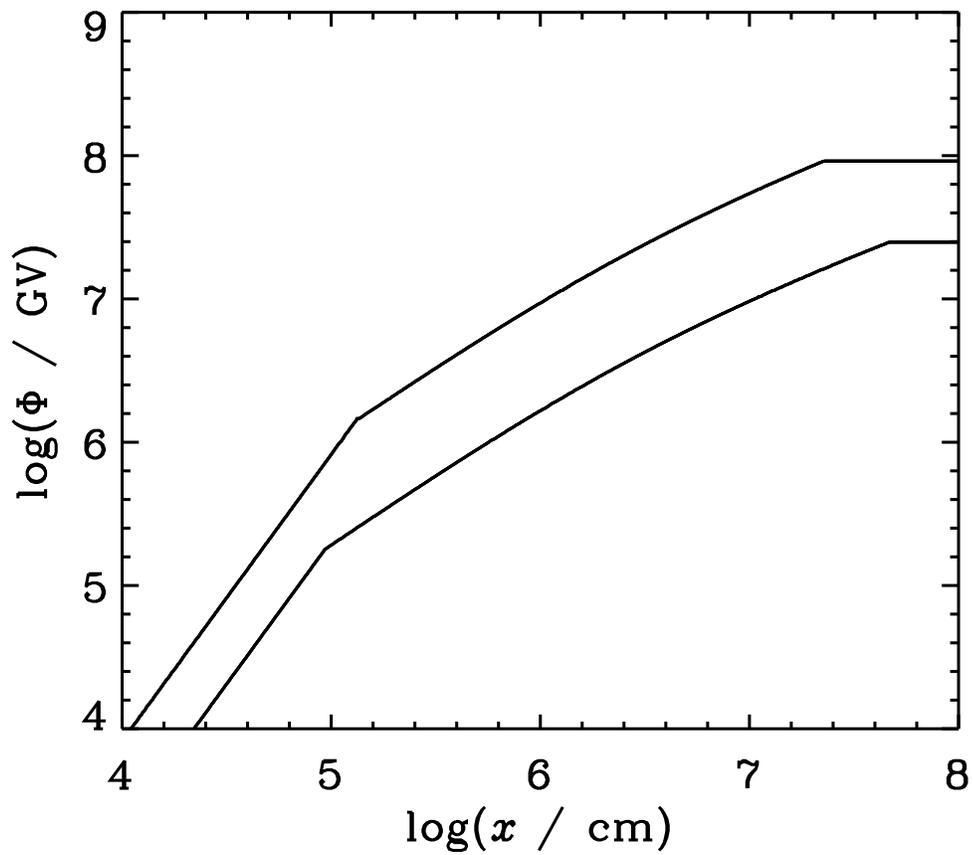}
\caption{Slot gap potential given by
Eq.~\protect\ref{Eq:slot_gap} for $B \sin i = 10^{12}$~G and $P=10$~ms 
(lower curve) and $P=5$~ms (upper curve). The
potential is assumed constant for $x$ larger than the light
cylinder radius. }
\label{fig:slot_gap_potential}
\end{figure}

\begin{figure}[htb]
\vspace{16cm}
\includegraphics{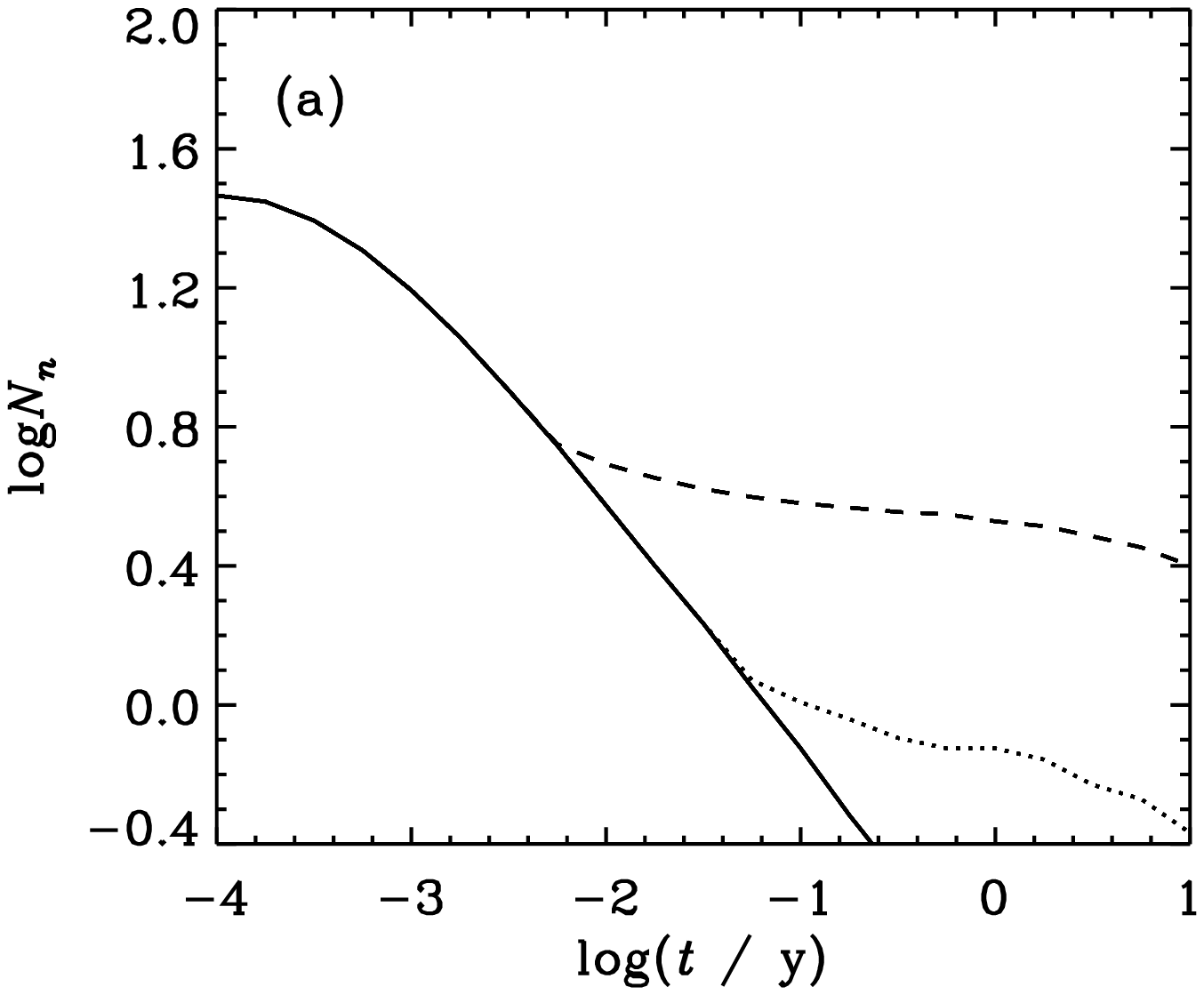}
\includegraphics{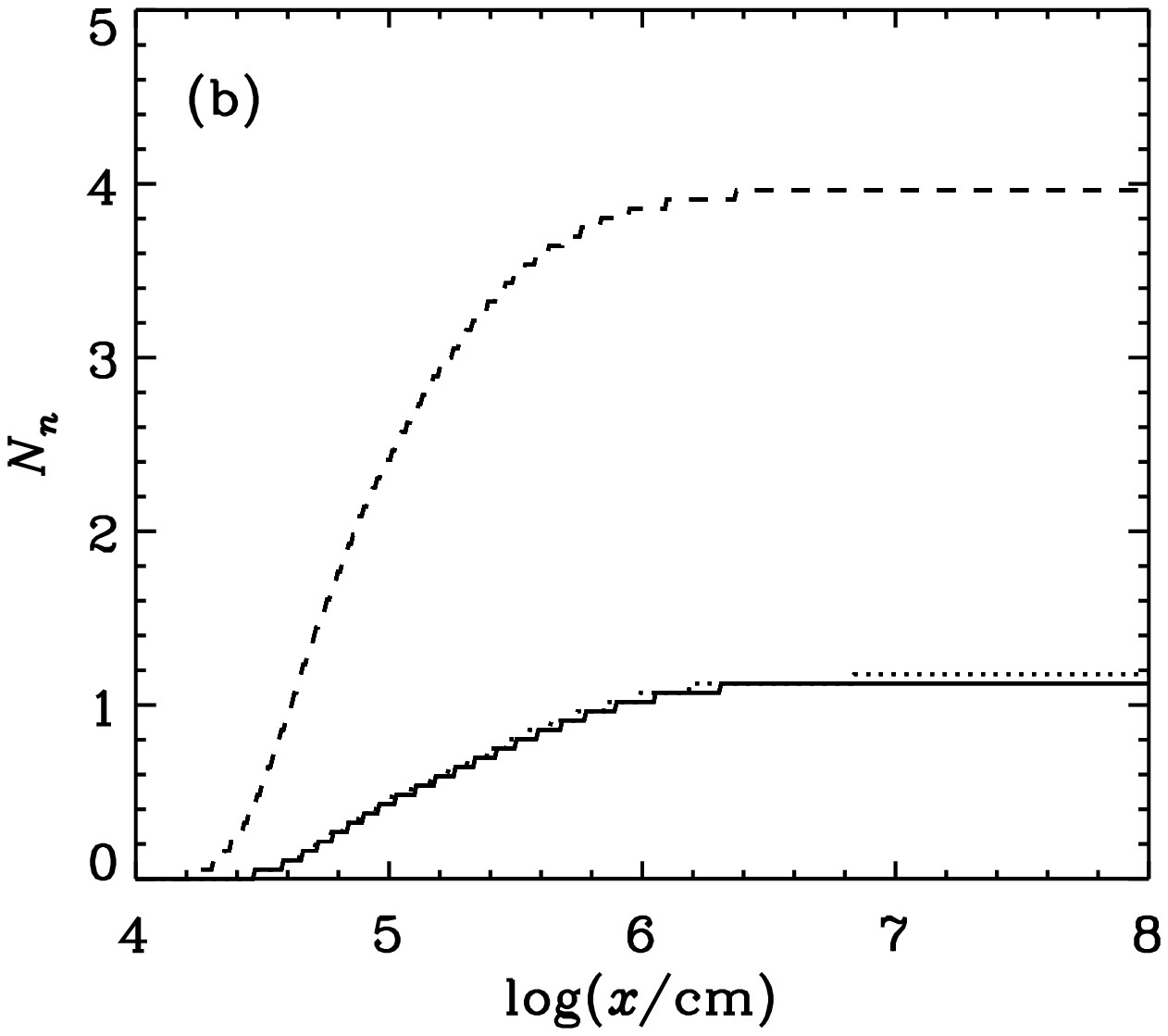}
\caption{Average cumulative number of neutrons extracted from a
single Fe-nucleus during acceleration by the assumed slot gap
potential for $B \sin i = 10^{12}$~G and $P_0=5$ ms through the
pulsar's radiation field: (a) as a function of neutron star
age; (b) as a function of distance along the slot
gap at $t=20.5$~days. Results are shown for no polar cap heating
(solid curve), moderate polar cap heating (dotted curve), and
maximum polar cap heating (dashed curve).  }
\label{fig:neutron_number}
\end{figure}

\begin{figure}[htb]
\vspace{16cm}
\includegraphics{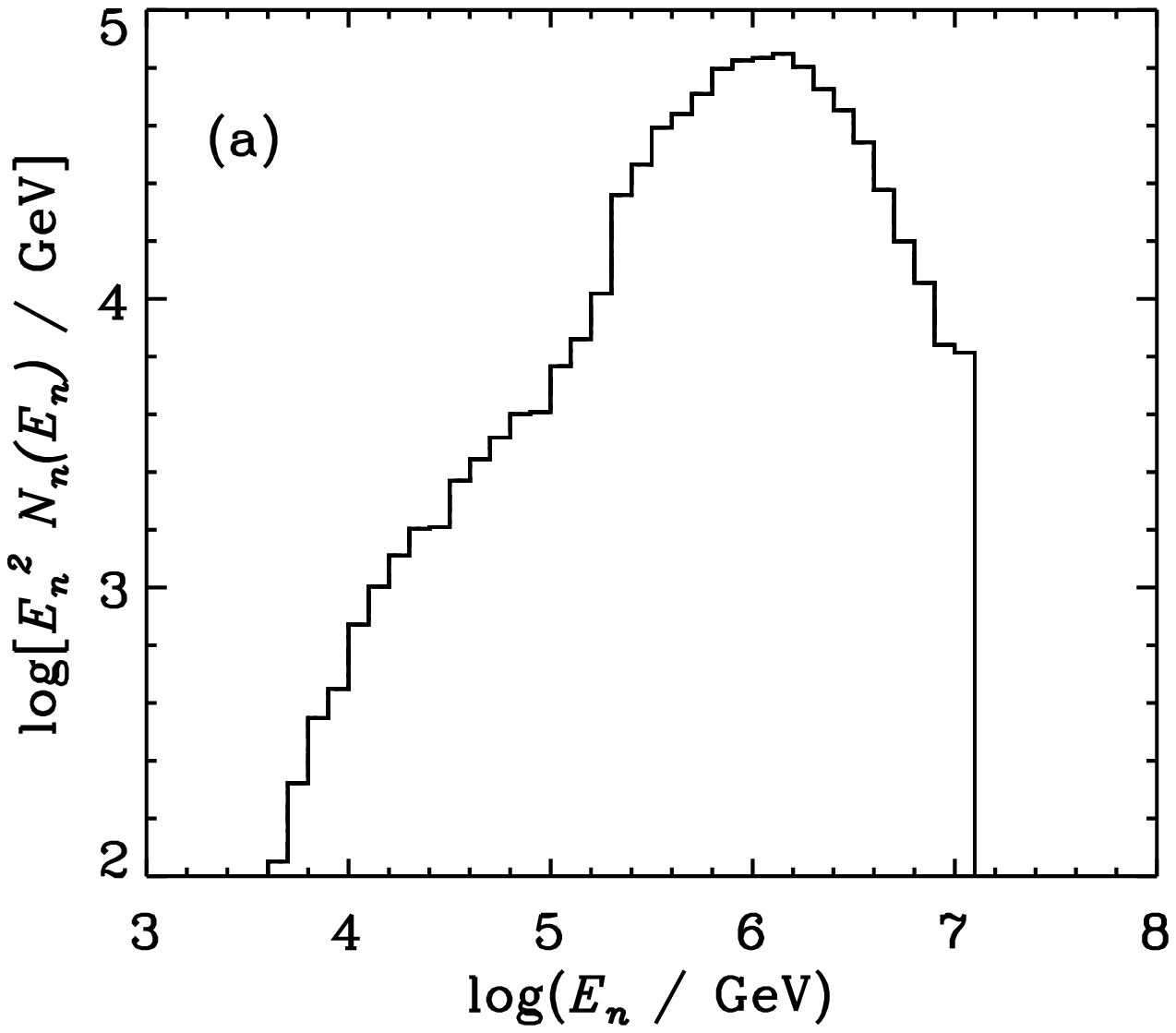}
\includegraphics{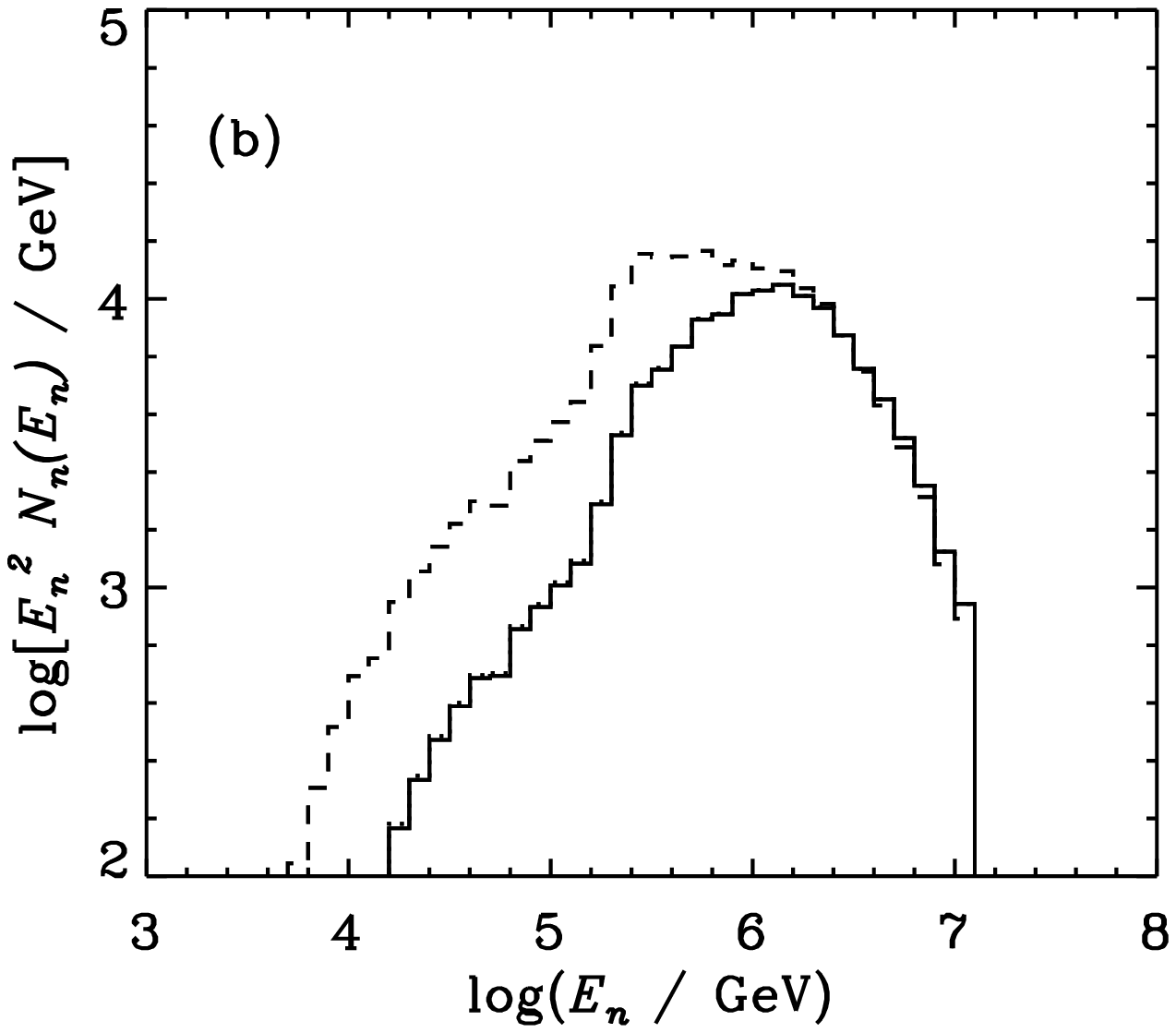}
\includegraphics{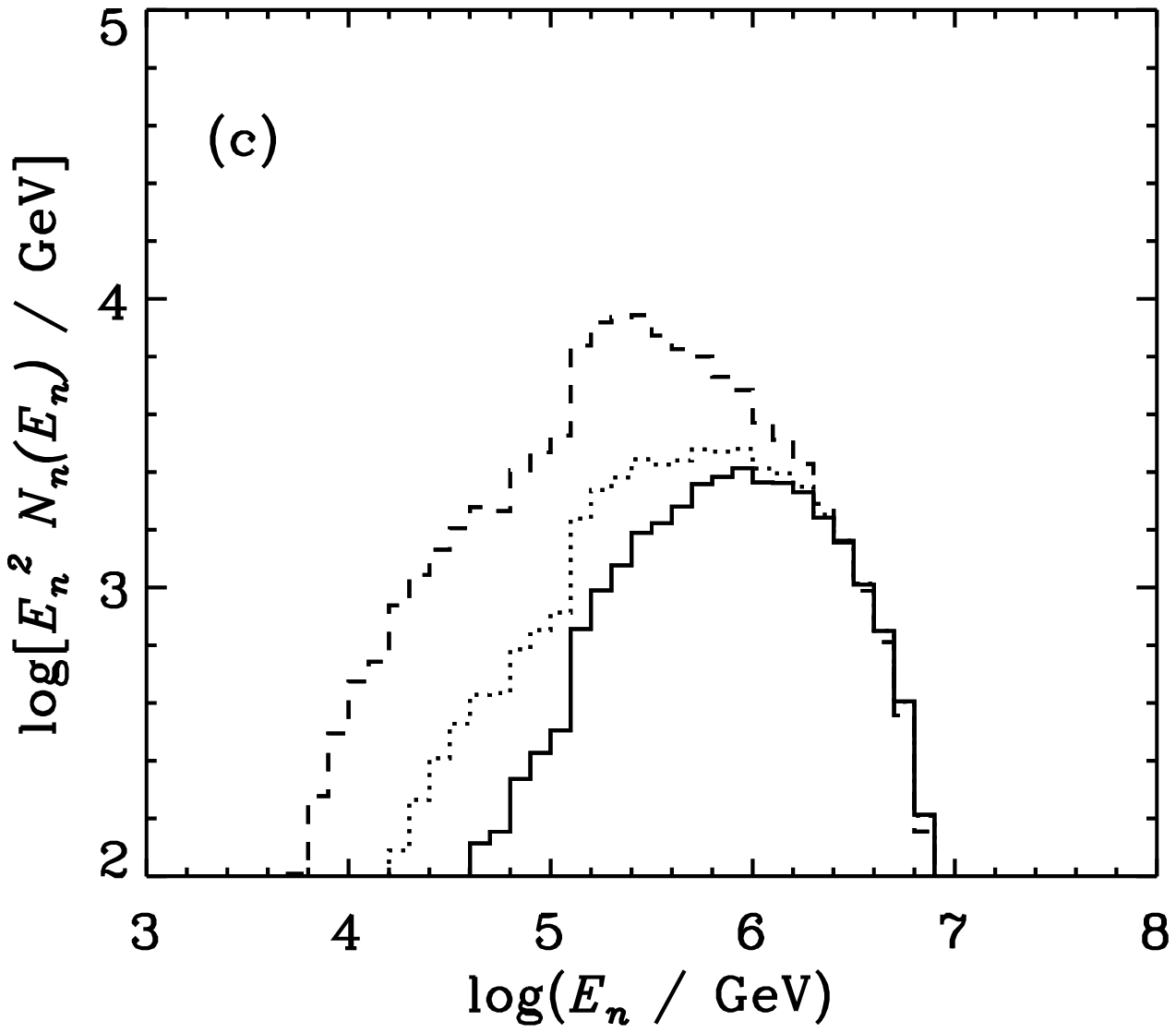}
\caption{The spectrum of neutrons extracted from a single
Fe-nucleus during acceleration for a pulsar with $P_0=5$~ms and
$B \sin i = 10^{12}$~G at (a) $t=1.15$~days ($P=5.00$~ms), (b) $t=20.5$~days
($P=5.05$~ms), and (c) $t=1$~year ($P=5.69$~ms).  Results are shown for
the three radiation field models: no polar cap heating -- solid
histogram; moderate heating -- dotted histogram; maximum heating
-- dashed histogram. }
\label{fig:neutron_spectra}
\end{figure}

\begin{figure}[htb]
\vspace{16cm}
\includegraphics{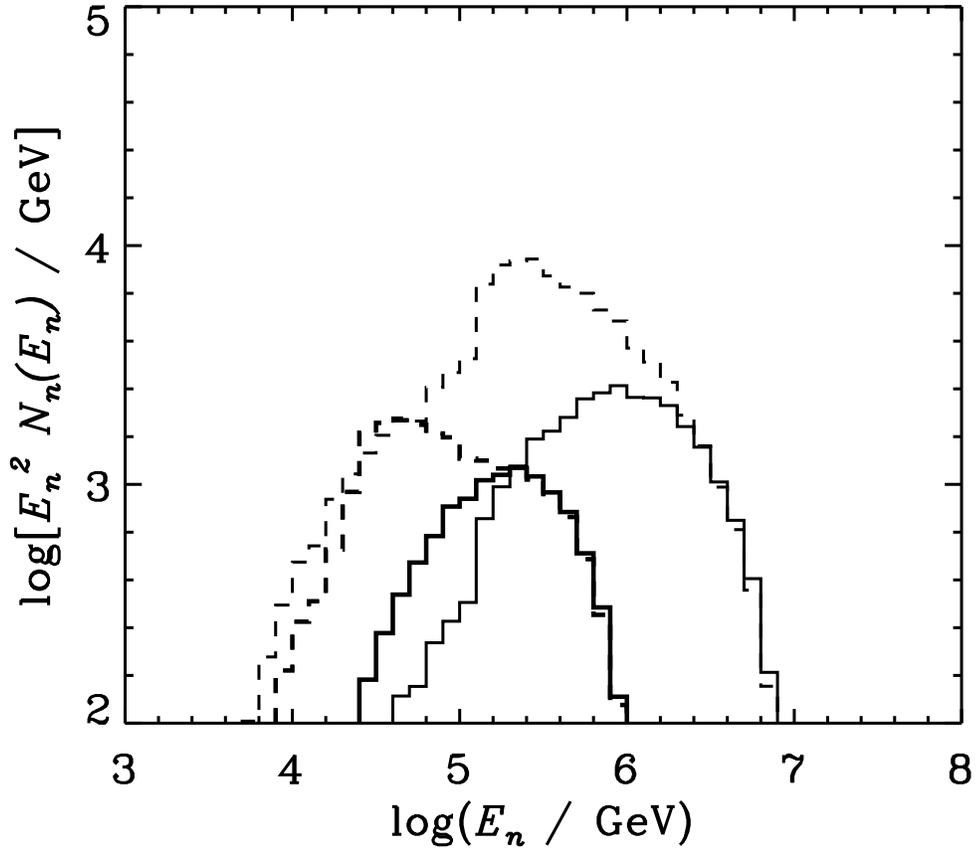}
\caption{The spectrum of neutrons extracted from a single
Fe-nucleus during acceleration at $t=1$ year and $B \sin i = 10^{12}$~G
for the following initial pulsar periods $P_0=5$~ms (thin histograms),
and $P_0=10$~ms (thick histograms).  Results are shown for two of the 
radiation field models: no polar cap heating -- solid histograms;
maximum heating -- dashed histograms. }
\label{fig:neutron_spectra_compare}
\end{figure}

\begin{figure}[htb]
\vspace{16cm}
\includegraphics{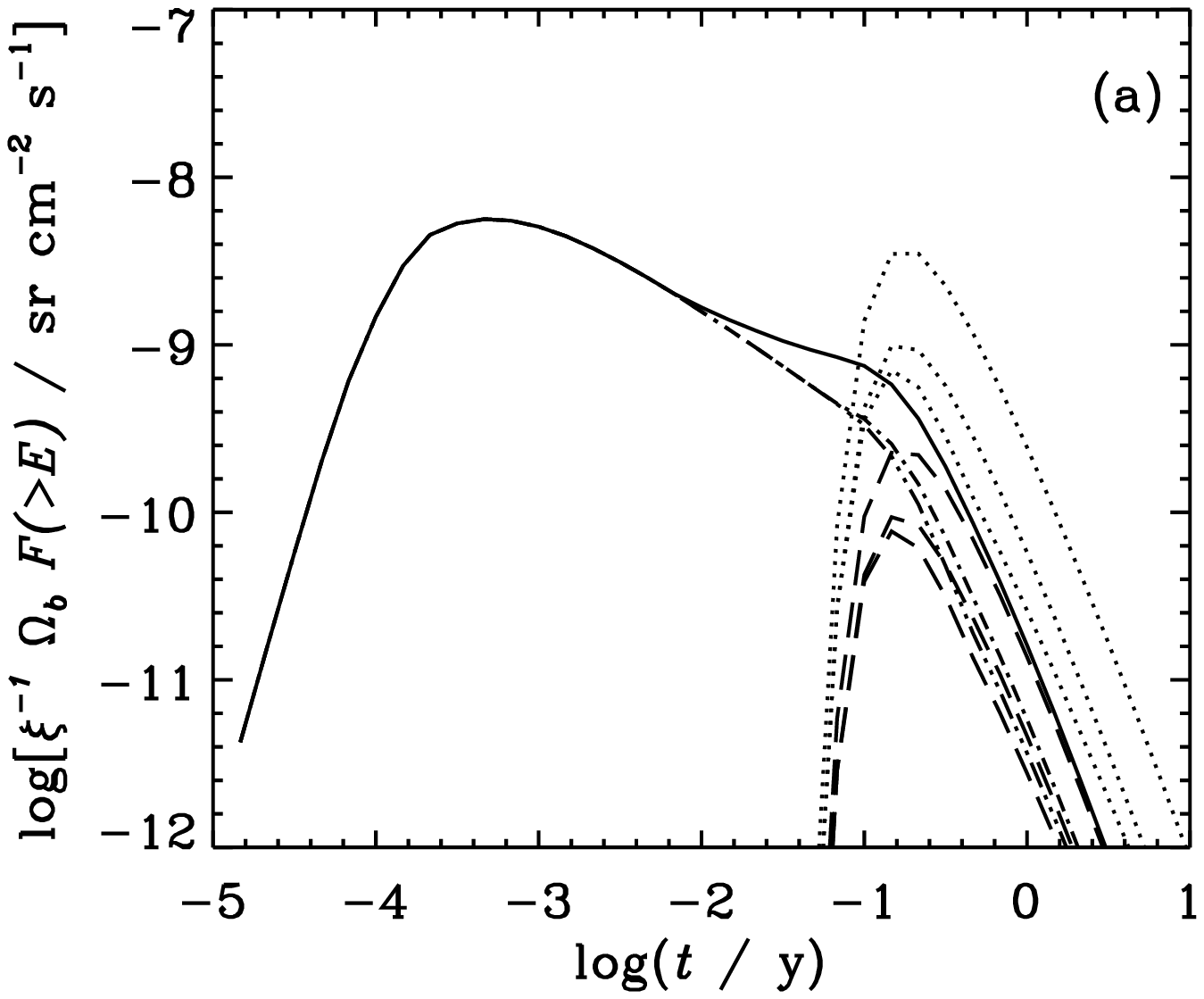}
\includegraphics{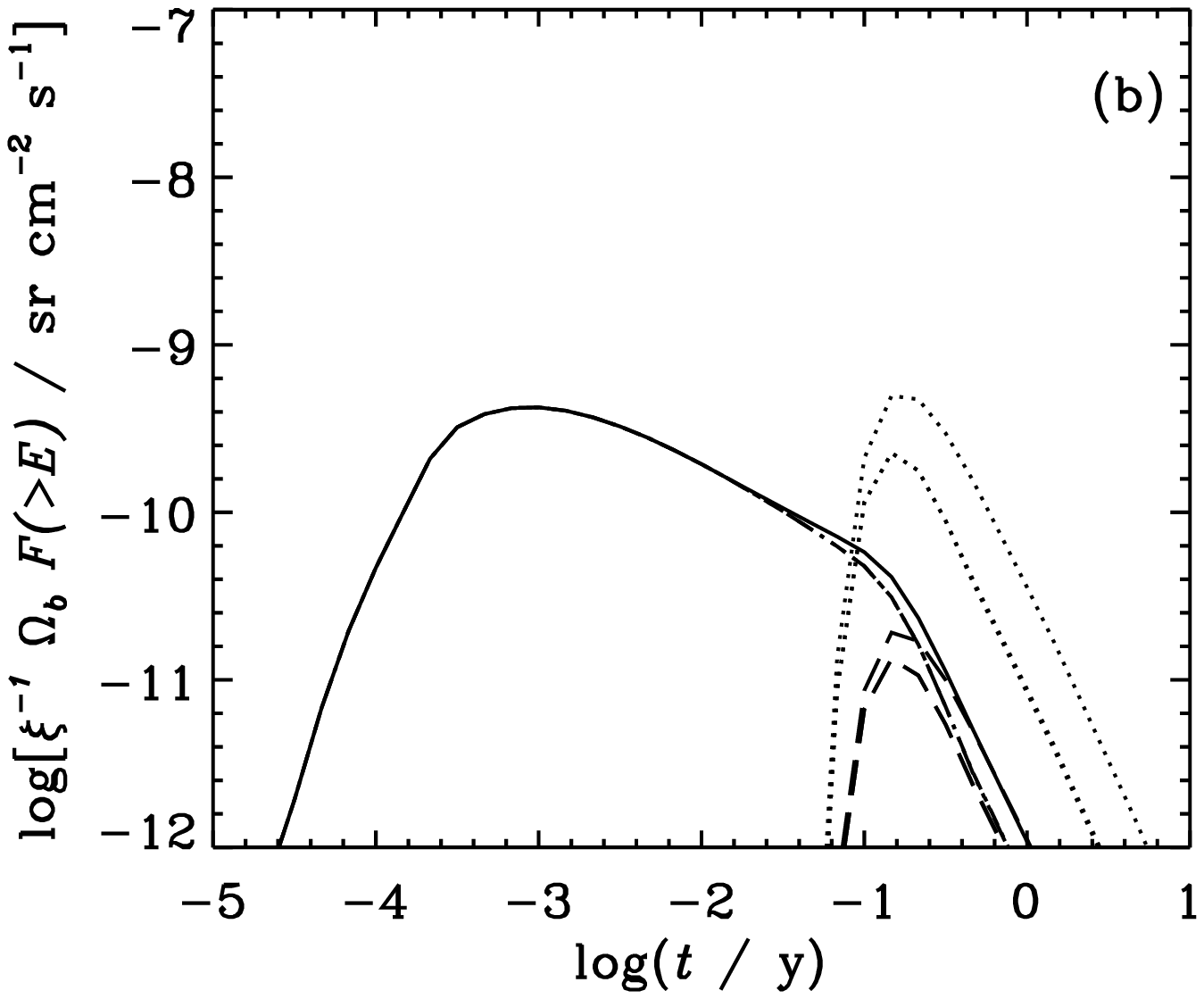}
\caption{The dependence of the integral flux of neutrinos $(\nu_\mu +
\bar{\nu}_\mu)$ above 1~TeV (solid, dot-dash, and
dot-dot-dot-dash curves), $\gamma$-rays above 100~MeV (dotted
curves), and $\gamma$-rays above 1~TeV (dashed curves) as a
function of time $t$ measured from the explosion for $d=10$~kpc
and $B \sin i = 10^{12}$~G and for (a) $P_0=5$~ms, and (b)
$P_0=10$~ms.  Results are shown for the three radiation field
models: no polar cap heating -- dot-dot-dot-dash curve and lowest
dotted and dashed curves; moderate heating -- dot-dash curve and
middle dotted and dashed curves; maximum heating -- solid curve
and upper dotted and dashed curves.  }
\label{fig:prompt_lightcurve}
\end{figure}

\begin{figure}[htb]
\vspace{16cm}
\includegraphics{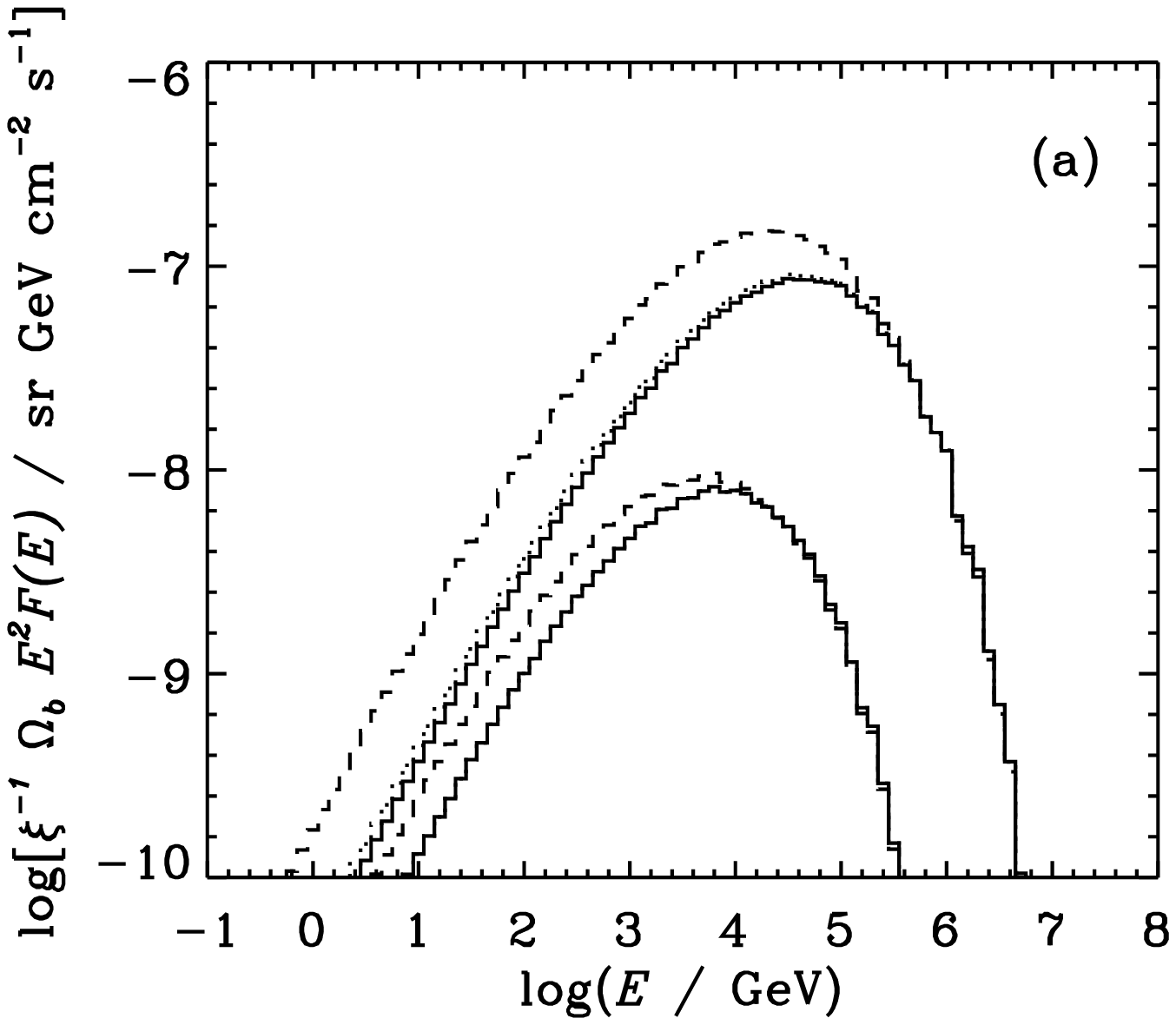}
\includegraphics{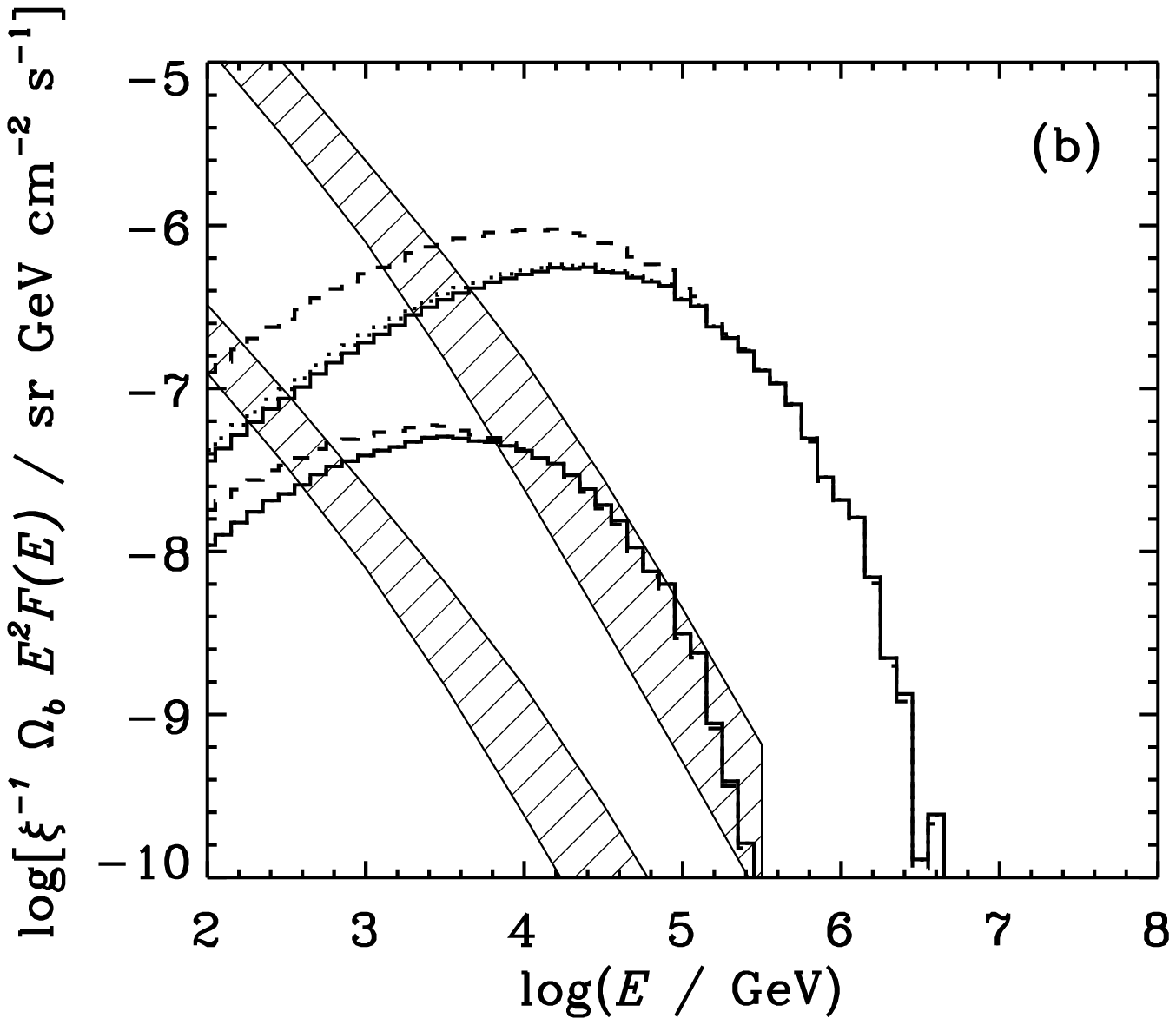}
\caption{The spectra for $d=10$~kpc at $t=0.1$ year of (a) 
$\gamma$-rays and (b)  neutrinos $(\nu_\mu +
\bar{\nu}_\mu)$ produced by collisions of neutrons with the
matter of the supernova shell. In each case, the upper three
histograms are for $P_0=5$~ms, and the lower three histograms for
$P_0=10$~ms.  Results are shown for $B \sin i = 10^{12}$~G for the three
radiation field models: no polar cap heating -- lower histogram;
moderate heating -- middle histogram; maximum heating -- upper
histogram.  The atmospheric neutrino background flux multiplied
by $E^2$ (GeV cm$^{-2}$ s$^{-1}$) within $1^\circ$ and within
$10^\circ$ of the source direction, based on the intensity
calculated by Lipari \protect\cite{Lipari}, is shown by the
hatched bands (each band shows the range of atmosheric neutrino
background as the zenith angle changes). }
\label{fig:prompt_spectrum}
\end{figure}

\begin{figure}[htb]
\vspace{16cm}
\includegraphics{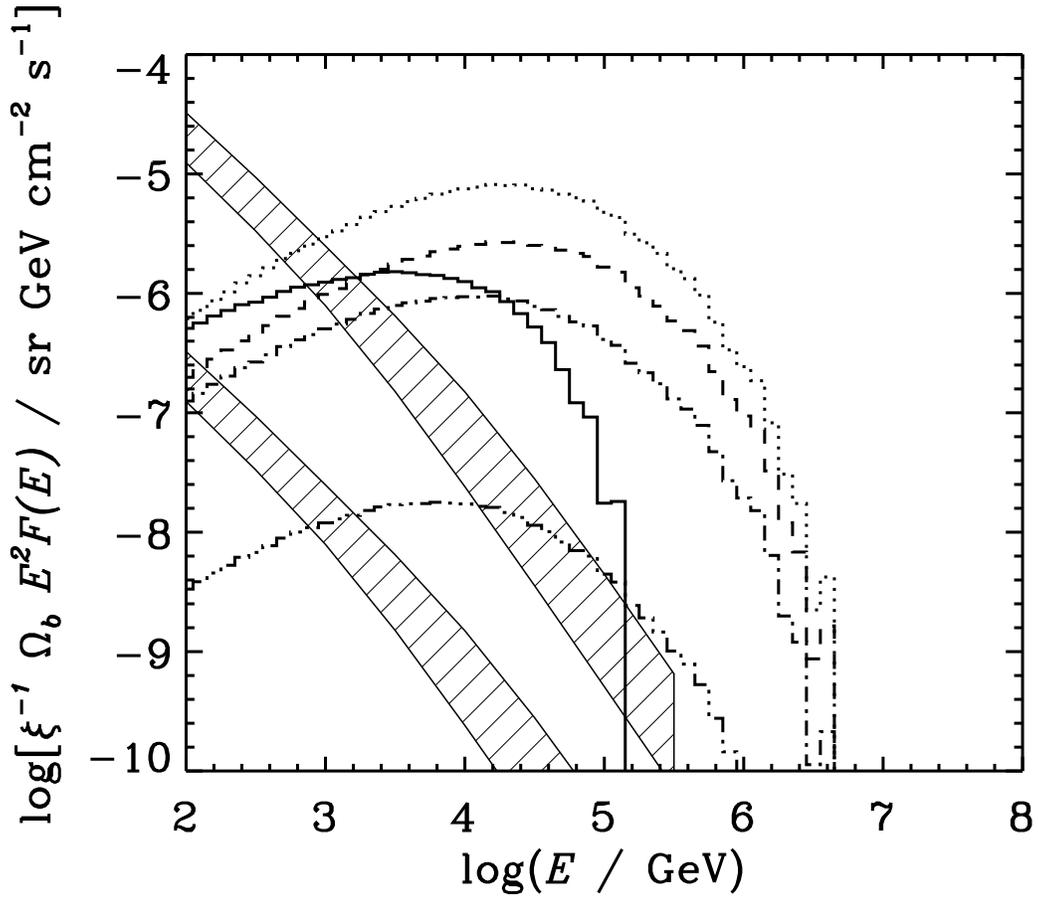}
\caption{Evolution of the spectrum of  neutrinos $(\nu_\mu
+ \bar{\nu}_\mu)$ produced by collisions of neutrons with 
matter in the supernova shell for $B \sin i = 10^{12}$~G, $P_0=5$~ms and the
maximum polar cap heating model.  Spectra are shown for
$d=10$~kpc and for $t=10^{-4}$~y (solid histogram); $10^{-3}$~y
(dotted histogram); $10^{-2}$~y (dashed histogram); $10^{-1}$~y
(dot-dash histogram); 1 y (dot-dot-dot-dash histogram).  See
Fig.~\protect\ref{fig:prompt_spectrum} for explanation of hatched bands.}
\label{fig:prompt_spectrum_ev}
\end{figure}

\end{document}